\documentclass{article}
\usepackage{amssymb,color,graphicx}
\usepackage{amssymb}
\usepackage{wrapfig,framed,cancel}
\usepackage[colorlinks=true, pdfstartview=FitV, linkcolor=blue, citecolor=black, urlcolor=blue]{hyperref}
\definecolor{shadecolor}{rgb}{0.94, 0.94, 0.87}
\def\ss{\mathbf s}
\newtheorem{theorem}{Theorem}[section]
\newtheorem{prop}{Proposition}[section]

\newtheorem{defn}{Definition}[section]
\newtheorem{lemma}{Lemma}[section] 

\newtheorem{cor}{Corollary}[section]
\newtheorem{example}{Example}[section]

\newtheorem{problem}{Problem}[section]
\newtheorem{remark}{Remark}[section]
\def\L{{\mathbf L}}
\def\ot{\hbox{\resizebox{6pt}{!}{$ \otimes \,
$}}}

\def\H{\mathcal H}
\def\r{ {\mathbf r}}
\def\A{{\mathbf A}}
\def \B{{\mathbf B}}
\def \C{{\mathbf C}}
\def \D{{\mathbf D}}
\def \K{{\mathcal K}}
\def\b{\beta}
\def\br{\begin{remark}}
\def\er{\end{remark}}

\def\s{\sigma}
\def\C{{\mathbb C}}

\def\eqref#1{ (\ref{#1})}
\def\&{&\hspace{-20pt}}
\def\R{{\mathbb R}}
\def\d{{\rm d}}

\def\1{\mathbf 1}
\def\a{{\alpha}}

\def\wh{\widehat}

\def\wt{\widetilde}
\def \pa{\partial}

\def\bea{\begin{eqnarray}}
\def\eea{\end{eqnarray}}
\def\Id{\mathrm{Id}}
\def\Ai{\mathrm{Ai}}

\def\res{\mathop{{\rm res}}}
\def\Tr{\mathop{{\rm Tr}}}
\def\Ai{\mathrm{Ai}}
\def\T{\mathrm{T}}

\def\Mat{\mathrm{Mat}}
\def\0{{\mathbf 0}}

\def\ov{\overline}
\def\QED{ {\bf Q.E.D}\par \vskip 4pt}
\newcommand{\be}{\begin{eqnarray}}
\newcommand{\ee}{\end{eqnarray}}
\newcommand{\bes}{\begin{eqnarray*}}
\newcommand{\ees}{\end{eqnarray*}}
\newcommand{\ds}{\displaystyle}

\definecolor{light-blue}{rgb}{0.8,0.85,1}
\definecolor{blue}{rgb}{0,0,1}
\definecolor{red}{rgb}{1,0,0}
\def\red#1{\textcolor[rgb]{0.9, 0, 0}{#1} }

\def\le{\left}
\def\ri{\right}
\def\ba{\begin{eqnarray}}
\def\eeq{\end{eqnarray}}

\makeatletter
\@addtoreset{equation}{section}
\makeatother

\begin{document}

\baselineskip 16pt plus 1pt minus 1pt
\begin{titlepage}
\begin{flushright}
\end{flushright}
\vspace{0.2cm}
\begin{center}
\begin{Large}
\textbf{Fredholm determinants and pole-free solutions to the noncommutative Painlev\'e II equation}
\end{Large}

\bigskip
M. Bertola$^{\dagger\ddagger}$\footnote{Work supported in part by the Natural
  Sciences and Engineering Research Council of Canada (NSERC)}\footnote{bertola@crm.umontreal.ca},  
M. Cafasso$^{\dagger\ddagger}$ \footnote{cafasso@crm.umontreal.ca}
\\
\bigskip
\begin{small}
$^{\dagger}$ {\em Centre de recherches math\'ematiques,
Universit\'e de Montr\'eal\\ C.~P.~6128, succ. centre ville, Montr\'eal,
Qu\'ebec, Canada H3C 3J7} \\
\smallskip
$^{\ddagger}$ {\em  Department of Mathematics and
Statistics, Concordia University\\ 1455 de Maisonneuve W., Montr\'eal, Qu\'ebec,
Canada H3G 1M8} 
\\
\end{small}
\end{center}
\bigskip
\begin{center}
\begin{abstract}
	We extend the formalism of integrable operators \`a la Its-Izergin-Korepin-Slavnov
 to matrix-valued convolution operators on a semi--infinite interval and to matrix integral operators with a  kernel  of the form  $\frac {E_1^T(\lambda) E_2(\mu)}{\lambda+\mu}$  thus proving that their resolvent operators can be expressed in terms of solutions of some specific Riemann-Hilbert problems.  We also describe some applications, mainly to a  noncommutative version of  Painlev\'e\ II (recently introduced by Retakh and Rubtsov), a related noncommutative equation of Painlev\'e\ type. We construct a particular family of solutions of the noncommutative Painlev\'e\ II that are pole-free (for real values of the variables) and hence analogous to the Hastings-McLeod solution of (commutative) Painlev\'e\  II. Such a solution plays the same role as its commutative counterpart relative to  the Tracy--Widom theorem, but  for  the computation of the Fredholm determinant of a matrix version of the Airy kernel.
\end{abstract}
\end{center}
\end{titlepage}
\tableofcontents
\section{Introduction and results}
The paper aims at extending the general theory of integrable operators of Its-Izergin-Korepin-Slavnov (IIKS for short)  \cite{IIKS} to operators of ``Hankel'' form (see below).
Leaving aside for the time being any analytical consideration, the issue is the study of integral operators on $L^2(\gamma_+,\C^r)$  with a kernel of the following form 
\be
\K(\lambda,\mu) = \frac {E_1^T(\lambda)E_2(\mu)}{\lambda + \mu}\label{introK} \ , \qquad \lambda,\mu\in \gamma_+
\ee
where $\gamma_+$ is a contour contained in a half-plane of $\C$ (so that the denominator does not vanish)\footnote{ The contour could be -for example- $\R_+$ provided that the matrices in the numerator yield $E_1^T(0)E_2(0)=0$.} and the matrices $E_j:\gamma_+ \to \Mat(p\times r,\C)$ are suitable (analytic) functions.
These operators are related via a Fourier transform to (matrix) convolution operators on $\R_+$ as pointed out in Section \ref{convolution}: our primary focus shall be the construction of a suitable Riemann--Hilbert problem for computing the resolvent operator $\mathcal S = -\K\circ (1+\K)^{-1}$.  The knowledge of the resolvent operator allows to write variational formul\ae\ for the Fredholm determinant of the operator $\Id + \K:L^2(\gamma_+, \C^r) \to L^2(\gamma_+,\C^r)$ via the well--known variational formula
\be
\pa  \ln \det \le(\Id + \K\ri) = \Tr \le((\Id + \mathcal S\ri) \circ \pa \K)\ .
\label{varform}
\ee
The situation is closely related to the IIKS theory mentioned above (with the tensorial extension explained in \cite{ItsHarnad}), which we briefly recall: let $\Sigma\subset \C$ be a collection of (smooth) contours and let ${\bf f}, {\bf g}:\Sigma \to \Mat( q\times n,\C)$ be smooth (analytic) functions on $\gamma$ subject to the condition  
\be
{\bf f}^T(\lambda) {\bf g}(\lambda) \equiv 0\ ,\ \ \lambda \in \Sigma\ .\label{nonsingular}
\ee
Consider the integral operator $N: L^2(\Sigma, \C^n)$ with kernel given by 
\be
N(\lambda,\mu):= \frac { {\bf f}^T(\lambda) {\bf g}(\mu)}{\lambda - \mu} \label{introIIKS}
\ee
Then the resolvent operator $\mathcal R = N\circ(\Id-N)^{-1}$ has a kernel (denoted with the same symbol $\mathcal R$) of the form\footnote{The superscript $^{-T}$ to a matrix denotes the inverse transposed matrix.} 
\be
\mathcal R(\lambda, \mu)  = \frac { {\bf f}^T(\lambda) \Theta^T(\lambda) \Theta^{-T}(\mu) {\bf g}(\mu)} {\lambda - \mu} 
\ee
where $\Theta(\lambda)$ is the  $q\times q$ matrix bounded solution of the following Riemann--Hilbert problem 
\be
\Theta(\lambda)_+ &\&= \Theta(\lambda)_- \le(\1_q - 2i\pi {\bf f}(\lambda) {\bf g}^T(\lambda)\ri)\nonumber \\
\Theta(\lambda)  &\& = \1_q + \mathcal O(\lambda^{-1}) \ ,\ \ \ \lambda \to \infty \label{RHPtheta}
\ee
Furthermore the solution of the RHP (\ref{RHPtheta}) exists if and only if the Fredholm determinant  $\det (\Id - N)$ is not zero.

The operator (\ref{introK}) is not immediately of the form (\ref{introIIKS}) and hence the IIKS theory is not directly applicable. Nevertheless the former situation is amenable -not surprisingly- to the latter (see also \cite{TWmKdV}). In fact one could observe, for example, that $\K^2$ is an operator of the form (\ref{introIIKS}) 
\be
\mathcal K^2(\lambda,\mu) &\& = \frac 1{\lambda-\mu} \le( E_1^T(\lambda) H_1(\mu) - H_2(\lambda)E_2(\mu)\ri)\nonumber \\
&\& H_1(\mu):= \int_{\gamma_+} \!\!\!\d\xi\frac {E_2(\xi)E_1^T(\xi)E_2(\mu)}{\mu+\xi} \ ,\ \ \ H_2(\lambda):=\int_{\gamma_+}\!\!\!\d\xi \frac {E_1^T(\lambda)E_2(\xi)E_1^T(\xi)}{\xi+\lambda}
\label{introK2}
\ee
This observation shows that the IIKS theory is relevant also to the study of operators of the form (\ref{introK}): however it is not practical to use (\ref{introK2}) as a starting point for the analysis as this route is impervious and is not the one we follow.
We provide a direct treatment of $\K$ as well as $\K^2$ in a unified fashion; the RHPs that  are relevant  are specified in Problems \ref{prob3}, \ref{prob1} (please refer to the statements there) for two matrix functions $\Gamma,\Xi$ of size $2r\times 2r$. The two problems are intimately related to each other in that the jump conditions are identical while only the asymptotic behavior at $\lambda=\infty$ for $\Gamma, \Xi$ differs.
 The solubility of the Riemann--Hilbert Problems \ref{prob3}, \ref{prob1} is equivalent to the non vanishing of the Fredholm determinants of the operators  $\Id_{_{\gamma_+}} - \K^2 $ (Thm. \ref{thmXi}) and $\Id_{_{\gamma_+}}+\K$ (Thm. \ref{thmGamma}), respectively,  which follows from IIKS theory; we thus obtain the formula for the resolvents of $\K$ and $\K^2$ (Theorems \ref{thmGamma} and \ref{thmXi}) 
\be
-\K\circ(\Id_{\gamma_+} + \K)^{-1}(\lambda,\mu)  =
\frac {2 \mu \le[E_1^T(\lambda),\0_{p\times r} \ri] \Gamma^{T}(\lambda) \Gamma^{-T}(\mu) 
\le[\begin{array}{c}
\0_{r\times p} 
\\
E_2(\mu)
\end{array} \ri]}{\lambda^2  - \mu^2}\\
 \K^2\circ(\Id_{\gamma_+} - \K^2)^{-1} (\lambda,\mu) = 
 [
E_1^T(\lambda)
,
\0_{p\times r}
] 
\frac{\Xi^{T}(\lambda)\Xi^{-T}(\mu)}{\lambda - \mu}\le[
\begin{array}{c}
\0_{r\times p}
\\
\!\!\! E_2(\mu)\!\!\! 
\end{array}
\ri]
\ee

The knowledge of the resolvent operator allows to write variational formul\ae\ for the respective Fredholm determinants: however one may bypass formula (\ref{varform}) and write the variational formul\ae\ {\em directly in terms of the solution of the respective RHPs} (Thm. \ref{4.1} \ref{4.2}) using the ideas in \cite{BertolaIsoTau}
\be
\pa \ln \det (\Id_{\gamma_+} - \mathcal K^2) &\& = \int_{\gamma_+\cup \gamma_-} \Tr \le( \Xi_-^{-1} \Xi_-' \pa M M^{-1}
	\ri) \frac {\d\lambda}{2i\pi}  \\
 \pa \ln \det (\Id_{\gamma_+} + \mathcal K)
&\&  =\frac 1 2 \int_{\gamma_+\cup \gamma_-} \Tr \le( \Gamma_-^{-1} \Gamma_-' \pa M M^{-1}
	\ri) \frac {\d\lambda}{2i\pi}\\
&\&M(\lambda)  := \1_{2r} - 2i\pi E_1(\lambda) E_2(\lambda)^T \le(\chi_{_{\gamma_+}}\ot \s_++  \chi_{_{\gamma_-}}\ot \s_-\ri), \ \gamma_-:= -\gamma_+ 
\ee
where $\prime$ is derivative w.r.t. $\lambda$, $\pa$ denotes any variation of the symbols $E_j$ and $\sigma_+$ ($\sigma_-$) denotes the $2\times2$ matrix with just one non-zero entry on the upper right corner (lower left corner). 

In the second part of the paper we provide some applications to the study of matrix convolution operators; our example of choice is a matrix version of the (scalar) convolution operator by the Airy function \cite{FSGOE}
\be
\mathcal Ai_s: &\&L^2(\R_+)\to L^2(\R_+)\nonumber \\
&\& f(y)\mapsto (\mathcal Ai_s f)(x):= \int_{\R_+} \Ai(x+y+2s) f(y)\d y
\ee
The Fredholm determinant of the operator $\Id - \mathcal Ai_{s/2}$  is known to yield the Tracy-Widom gap distribution $F_1(s)$ for the GOE \cite{FSGOE} and --on the other hand- the Fredholm determinant of $\Id - \mathcal Ai_s^2$ yields the distribution $F_2(s)$ for the GUE \cite{MehtaBook} ; in fact it is well known that the kernel of the square of the Airy-convolution operator is the celebrated Airy kernel 
\be
\mathcal Ai^2(x,y) :=\int_{R_+} \Ai(x+z)\Ai(y+z) \d z = \frac {\Ai(x)\Ai'(y) - \Ai(y)\Ai'(x)}{x-y} =:K_{\Ai}(x,y)
\ee
 $F_2$ is expressed in terms of the Hastings-McLeod solution  \cite{Hastings-McLeod} to the second Painlev\'e\ equation \cite{TracyWidomLevel} while $F_1$ can be expressed in terms of the Miura transform of the same transcendent. Alternatively (and equivalently) $F_1(s)$ can be expressed in terms of the unique solution of the Painlev\'e XXXIV equation with a certain prescribed asymptotics\footnote{The uniqueness of the solution $w$ with the prescribed asymptotics is easily deduced from the uniqueness of the Hasting-Mc Leod solution $u$ of PII.}.
\bea
	F_2(s) &\&= \exp\le(-\int_{s}^\infty(x-s)u(x)^2dx\ri)\ ,\qquad u^2(s)  =-\pa_s^2 \ln F_2(s) \label{TW4}\\
	F_1(s) &\&= \exp\le(-\frac{1}{2}\int_{s}^\infty u(x)dx\ri)\Big(F_2(s)\Big)^{\frac{1}2}\label{TW3}\\
	F_1(s) &\&=\exp\le(-\int_{s}^\infty(x-s)w(x)dx\ri)\ , \qquad w(s) = -\pa_s^2\ln F_1(s) \label{TW3bis}\\
	u''(s)&\&= 2u(s)^3+su(s),\quad u(s)\sim\Ai(s),\quad s\longrightarrow +\infty.\label{TW5}\\
	w'''(s)&\&= 12w(s)w'(s)+2w(s) + s w'(s),\quad w(s)\sim-\frac{1}{2}\Ai'(s),\quad s \longrightarrow +\infty.\label{P34int}\\
	w(s)&\& = \frac 1 2 u^2(s) - \frac 1 2 u'(s)\label{Miuraint}
\eea
where (\ref{Miuraint}) is the usual Miura transformation between solutions of modified KdV and KdV equations. Moreover we refer to (\ref{P34int}) as the PXXXIV equation since, up to rescaling, this is the same as the derivative of equation (30) in \cite{CJP} (see also \cite{KajiwaraMasuda}).

The noncommutative analog of the whole preceding discussion arises in the study of a matrix version of the Airy-convolution operator (see Section \ref{ncAirykernel}) which we have picked as exemplary application:
\be
(\mathcal Ai_{\vec s}\,\vec f)(x)&\&  := \int_{\R_+} {\mathbf {Ai}}(x+y;\vec s) \vec f(y)\d y\\
{\bf Ai} (x;\vec s)&\& := \le[c_{jk} \Ai(x+s_j+s_k)\ri]_{j,k}\ .
 \ee
Here the matrix $C=[c_{jk}]_{j,k}$ is an arbitrary $r\times r$ matrix with complex entries (in general) and the dependence of $\mathcal Ai_{\vec s}$ on $C$ is considered as parametric (and it is understood in the notation). The kernel of the square of this matrix-kernel does define a probabilistic model because it is a totally positive kernel on the configuration space $\{1,\dots, r\}\times \R$ as shown in Thm. \ref{thmdetpp}.
The analysis of $\mathcal Ai_{\vec s}$ and $\mathcal Ai_{\vec s}^2$ is then related to certain noncommutative analogs of the aforementioned Painlev\'e\ equations and particular solutions thereof; in particular the Fredholm determinant of $\mathcal Ai_{\vec s}^2$  is related to the noncommutative (matrix) Painlev\'e\ II equation\footnote{Equation (\ref{introPII}) reduces to (\ref{TW5}) in the scalar case $r=1$ with the change of variable $x=2s$. Also, the matrix $U(\vec s)$ in the body of the paper shall be denoted by $\b_1(\vec s)$.}
\be
\D^2 U(\vec s) = 4 \le( \ss U (\vec s) + U(\vec s) \ss\ri) + 8  U^3(\vec s)\ ,\ \ \  \ss:= {\rm diag} \le(s_1,\dots, s_r\ri) \ ,\ \ \ \D:= \sum_{j=1}^r \frac \pa{\pa s_j}\ .\label{introPII}
\ee
which appeared recently in \cite{RetakhRubtsov}: that paper provided special solutions in terms of {\em quasideterminants}
\cite{Gelfand2RetakhWilson} in a more general context of noncommutative rings, but not a Lax-pair representation or a connection to Riemann--Hilbert problems or Fredholm determinants.
 The isomonodromic approach to the above equation yields a Lax pair representation contained in Section \ref{PIILax} and particularly Lemma \ref{nc1}.  

Of greater interest is the fact that the particular solution that is involved in the computation of the Fredholm determinant of $\mathcal Ai_{\vec s}^2$ enjoys the same smoothness properties for $\vec s\in \R^r$ as the Hastings-McLeod solution. More precisely we prove (Prop. \ref{HMunique}) that there is a unique solution of (\ref{introPII}) with the asymptotic
\be
U_{k\ell}(\vec s) = c_{k\ell} \Ai(s_k + s_\ell) + \mathcal O\le (\!\! \sqrt {S} {\rm e}^{-\frac 43 (2S-2m)^\frac 32}\!\!\ri) ,  \ S:=\frac 1r \sum s_j ,\ \ m=\max_j |s_j-S|\, , \ S\!\to\! +\infty\nonumber \ .
\ee
Additionally, this solution is {\bf pole free} for $\vec s\in \R^r$ if the maximal singular value of the matrix $C=[c_{k\ell}]$ is one or less\footnote{The singular values of a matrix $C$ are the (positive) squareroots of the eigenvalues of $C^\dagger C$: they coincide with the absolute values of the eigenvalues of $C$ if it is Hermitean (or more generally normal).}, a condition which is sufficient if $C$ is an arbitrary complex matrix and becomes also necessary if $C$ is Hermitean (Thm. \ref{Xiexists} and Thm. \ref{thmncHML}).
The analog of the third-order ODE for $F_1$ is now a system with noncommutative symbols (Thm. \ref{thm62}) that can be reduced to a {\em fourth} order matrix ODE (Remark \ref{rem61}) and only in the scalar case is further reduced to an ODE of the third order. The Fredholm determinant of $\mathcal Ai_{\vec s}$ (the analog of $F_1$) is then computed in terms of the relevant solution in Corollaries \ref{CorTW1}, \ref{CorTW2}.

\section{Matrix convolution operators on a semi--infinite interval.}\label{convolution}
Given a function ${\bf C}:\R\longrightarrow \Mat(r \times r)$ decaying sufficiently fast at infinity let's consider the convolution operator ${\mathcal C}$ acting on  $L^2(\R_+,\C^r)$ as follows:
\be
\big(\mathcal C\varphi\big)(x) = \int_0^\infty {\bf C}(x+y) \varphi(y)\d y\in L^2([0,\infty), \C^r)
\ee
Our aim is to study the Fredholm determinants $\det(\Id+
\mathcal C)$ and $\det(\Id-\mathcal C^2)$\footnote{Of course the sign in the expression $\det(\Id + \mathcal C_s)$ is inessential since we can always change ${\bf C}$ with $-{\bf C}$.}. Here $\mathcal C$ (and hence the determinant)  may depend on some parameters not explicitly indicated here (see below). Such type of determinants appears in many applications; just to cite two of them let's recall the Dyson formula  \cite{DysonFormula} in the inverse scattering for the Schr\"odinger operator and in the integral formula of the Tracy-Widom distribution for GOE found by Ferrari and Spohn \cite{FSGOE} (see below) for $r=1$. 
\br
In the inverse scattering theory of the Schr\"odinger operator and other applications the Fredholm determinant is written as the restriction to $[s,\infty)$ of the convolution by ${\bf C}$:
\be
f\mapsto \int_s^\infty {\bf C}(x+y)f(y)\d y \in L^2([s,\infty)).
\ee
 This is identical to the setting above up to translation. In fact it is just enough to redefine ${\bf C}(x)\longmapsto {\bf C_s}(x):={\bf C}(x+2s)$ and make it act on $L^2([0,\infty)$.
\er
We will consider functions ${\bf C}(z)$ that admit the following representation (the factor of $-i$  being purely for later convenience)
\be
{\bf C}(z) = -i \int_{\gamma_+} {\rm e}^{i z\mu } \r(\mu) \d \mu
\label{2.3}\ee
where $\gamma_+$ stands for a finite  union of   oriented contours in the upper-half plane with positive distance from $\R$ and $\r(\mu)$ is a bounded  $L^{1}(\gamma_+, \Mat(r\times r))$ function on $\gamma_+$ (with respect to the arc-length measure).
This assumption guarantees that ${\bf C}(z)$ is rapidly decaying at $z=+\infty\in \R$ with a simple estimate\footnote{The symbol $|\r |$ on a matrix stands for any norm on the matrices, for example the Hilbert-Schmidt norm or the supremum of the absolute values of the entries. This is so not to overload the notation when considering norms in some $L^p$.}
\be
|{\bf C}(z)| \leq {\rm e}^{-z {\rm dist} (\gamma_+,\R)} \int_{\gamma_+} |\r(\mu)| |\d\mu|.
\ee

An interesting example is as follows

\begin{example}
\label{Airy}
Let ${\bf C}(z) =-\Ai(z)$ and $r=1$: then 
\be
{\bf C}(z)=-\Ai(z+s/2) =-\frac 1{2\pi} \int_{\gamma_+} {\rm e}^{i \frac {\mu^3} 3 + i (z+s/2) \mu}\d\mu
\ee
where $\gamma_+$ is a contour extending to infinity along the directions $\arg(\mu) = \frac \pi 2  \pm \frac {\pi}3$. This example is relevant for applications since, as we have written in the introduction, the Fredholm determinant of the corresponding convolution operator is equal to the Tracy-Widom distribution for GOE, namely
$$F_1(s)=\det(\Id+\mathcal C).$$
\end{example}

We would like to transfer the study of the Fredholm determinant of $\mathcal C$ on $L^2(\R_+)$ to the study of a Fredholm determinant of an operator in  $L^2(\gamma_+)$; this is accomplished hereafter.
\begin{prop}\label{KConv}
	Let $C(z)$ as above, with $\r(\mu) =E_1(\mu) E_2^T(\mu)$ and $E_j \in L^2\cap L^\infty(\gamma_+, \Mat(r\times p))$; then the operator $\mathcal C$ is of trace--class on $L^2(\R_+,\C^r)$ and also  
\be
	\det(\Id_{L^2(\R_+)}
	+ \mathcal C)=\det(\Id_{\H^2_r}+\wh \K)
\ee
where $\H^2_r$ is the Hardy space $\H^2\otimes \C^r$ (i.e. the unitary image of the Fourier--Plancherel transform of $L^2(\R_+,\C^r)$) and 
where $\wh \K$ is the integral operator on $\H^2_r\subset L^2(\R,\C^r)$ with kernel 
\be
	\wh \K(\lambda,\mu):=\int_{\gamma_+} \frac {\r^T(\xi)\d\xi}{2i\pi(\lambda-\xi)(\mu+\xi)}\label{khatkern}
\ee 
\end{prop}
{\bf Proof of Prop. \ref{KConv}.}
 By Paley--Wiener theorem, $L^2(\R_+,\C^r)$ is unitarily equivalent under Fourier--Plancherel transform $\mathcal T$ to the subspace $\H^2_r := \H^2\otimes \C^r$, with $\H^2$ the Hardy space of the upper half plane.
Hence the convolution operator acts as follows 
\bea
\psi(x):= (\mathcal C\varphi)(x)
= \int_0^\infty C(x+y) \varphi(y)\d y 
= -i\int_0^\infty \!\!\!\! \d y  \int_{\gamma_+} \!\!\!\!\! \d\xi\,{\rm e}^{i(x+y)\xi} \r(\xi) \varphi(y) = \nonumber\\
=-i\sqrt{2\pi} \int_{\gamma_+} \!\!\!\!\!\! \d\xi\,{\rm e}^{ix\xi}\r(\xi) ({\cal T} \varphi)(\xi)\nonumber
\eea
so that (note that the $x$--integral below is convergent because $\xi \in \gamma_+\subset\C_+$)
\bea
({\cal T} \psi)(\lambda) 
= \frac 1{\sqrt{2\pi}}\int_0^\infty {\rm e}^{i\lambda x} \psi(x)\d x  
= 
-i\int_0^\infty\!\!\!\! \d x\,  {\rm e}^{i\lambda x}\int_{\gamma_+} \!\!\! \d\xi\, {\rm e}^{ix\xi} \r(\xi) ({\cal T} \varphi)(\xi)=\nonumber\\
= \int_{\gamma_+} \!\!\! \d\xi\,\frac{\r(\xi)}{\lambda+\xi} ({\cal T}\varphi)(\xi)
 = \int_{\gamma_+} \!\!\! \d\xi\,  \frac{\r(\xi)}{\lambda+\xi} ({\cal T}\varphi)(\xi)\ .
 \eea
We note that for a function in $\H^2_r$ like $ f(\mu):= {\cal T}\varphi(\mu)$,  the evaluation at a point $\xi\in \C_+$ can be written as 
 \bea
 f(\xi)  =\int_{\R} f(\mu) \frac{\d\mu}{2i\pi(\mu-\xi)} \ ,
\eea
which is Cauchy's theorem. We shall thus define 
\be
\wh \K^T := \mathcal T^{-1} \mathcal C\mathcal T
\ee
(the reason for the transposition is solely for later convenience)
with kernel given by
\be
\wh \K^T f(\lambda)  = \int_\R \d\mu \int_{\gamma_+} \!\!\! \d\xi\,  \frac{\r(\xi)}{\lambda+\xi} \frac {f(\mu)}{2i\pi(\mu-\xi)}\ \ \Rightarrow 
\wh \K f(\lambda)  = \int_\R \d\mu \int_{\gamma_+} \!\!\! \d\xi\,  \frac{\r^T(\xi)}{\lambda-\xi} \frac {f(\mu)}{2i\pi(\mu+ \xi)}\ \ 
\ee
Finally, since the Fourier Plancherel transform from $L^2(\R_+,\C^r)$ to $\H^2_r$ is an isometry, the respective Fredholm determinants are equal (if they exist).  We note that $\wh \K$ extends to an integral operator on the whole of $L^2(\R,\C^r)$ with the same kernel: this extension automatically annihilates the orthogonal complement of $\H^2_r$ in $L^2(\R,\C^r)$, which is seen by closing the $\mu$--integral with a half circle in the lower half plane and then invoking Cauchy's theorem. We will understand this extension in what follows.

Therefore to conclude we need to show that $\mathcal C$  and $\wh \K$ are trace-class. By the unitary equivalence given by the Fourier-Plancherel transform it suffices to show that $\wh \K$ is trace class:
we shall present $\wh \K$ as the product of two Hilbert-Schmidt operators, thus proving it of trace class.
To this end recall that $\r(\mu) = E_1(\mu)E_2^T(\mu)$; thus 
\be
\wh \K f (\lambda) =\int_\R\!\!\!\d\mu  \int_{\gamma_+} \!\!\! \d\xi\,  \frac{E_2(\xi)}{\lambda-\xi} \frac {E_1^T (\xi)f(\mu)\d\mu}{2i\pi(\mu+\xi)} = \mathcal C_2\circ \mathcal C_1 f(\lambda)\ .
\ee
where the two operators are defined as follows
\be
\mathcal C_1&\&:\H^2_r \subset L^2(\R,\C^r) \to L^2(\gamma_+,\C^p)\nonumber \\
&\& f\mapsto (\mathcal C_1 f)(\xi):= E_1^T(\xi) \int_{\R} \frac {f(\mu)\d\mu}{2i\pi(\mu + \xi)}\label{C1}
\ee
\be
\mathcal C_2&\& : L^2(\gamma_+,\C^p) \to \H_r^2 \nonumber \\
&\& h\mapsto(\mathcal C_2 h)(\lambda):=  \int_{\gamma_+} \frac {E_2(\xi) h(\xi)\d\xi}{-\xi+\lambda} \label{C2}
\ee
We embed $\H^2_r$ and $L^2(\gamma_+,\C^p)$ as subspaces of $L^2(\R\cup \gamma_+,\C^{r+p})$
as
\be
\H^2_r\ni f \mapsto \le[\begin{array}{c}
f(\lambda)\chi_{_\R}(\lambda)\\ \vec 0_p
\end{array}\ri]\ ,\qquad L^2(\gamma_+,\C^p) \ni h \mapsto \le[\begin{array}{c}
\vec 0_r\\
h(\lambda)\chi_{_{\gamma_+} }(\lambda)
\end{array}\ri]
\ee
 (they are then orthogonal but not complementary) and think of $\mathcal C_j$ as extended to the whole $L^2(\R\cup \gamma_+,\C^{r+p})$ in the trivial way (i.e. acting like zero on the orthogonal complements of the $\H^2_r$, $L^2(\gamma_+, \C^p)$, respectively). Analogously we extend trivially the action of $\wh K$ to this enlarged Hilbert space.
Then it is promptly seen that they both are Hilbert Schmidt  in $L^2(\R\cup \gamma_+,\C^{r+p})$ because
\be
\int_{\gamma_+}\!\!\!\! |\d\xi|\int_{\R}\!\!\!  |\d \mu|\,\frac{\Tr\le( E_j^\dagger(\xi) E_j(\xi)\ri)}{|\xi \pm \mu|^2}<+\infty
\ee
thanks to our assumption that (the entries of) $E_j$ are all in $L^2(\gamma_+,\C)$. Thus $\wh {\mathcal K}: \H^2_r\to \H^2_r$ is trace class, so is $\mathcal C$ and their determinants are the same. \QED 

Recall that  if $A: \H_1\to \H_2$ and $B:\H_2\to \H_1$ are (bounded) operators between Hilbert spaces  and both $AB$, $BA$ are trace class then (see for instance \cite{Si})
\be
\Tr_{\H_1} (B\circ A) = \Tr_{\H_2} (A \circ B)\ .
\ee
Composing the operators $\mathcal C_j$ in the opposite order we obtain an operator on $L^2(\gamma_+,\C^p)$ as follows

\bea
(\mathcal C_1\circ \mathcal C_2 f)(\mu)=\frac { E_1^T(\mu)}{2i\pi}  \int_{\R} \d \xi \int_{\gamma_+} \d\lambda \frac {E_2(\lambda)f(\lambda)}{(\xi - \lambda)(\xi + \mu)} 
\eea
The $\xi$--integral can be closed with a big circle in the upper--half plane, thus picking up the residue at $\xi=\lambda$ to give 
\be
(\mathcal C_1\circ \mathcal C_2 f)(\mu)={ E_1^T(\mu)}\int_{\gamma_+} \d\lambda \frac {E_2(\lambda)f(\lambda)}{\lambda + \mu} =:({\mathcal K} f)(\mu).
\ee
Renaming the variables we obtain that
\be
{\mathcal K}(\lambda,\mu) = \frac {E_1^T(\lambda) E_2(\mu)}{\lambda + \mu}\ . \label{kernel}
\ee
The operator $\mathcal K$ is also trace class because the composition of two Hilbert-Schmidt  operators in $L^2(\gamma_+\cup \R, \C^{r+p})$; clearly it defines an operator on $L^2(\gamma_+,\C^p)$ since it acts trivially on its orthogonal complement (by construction).
In particular 
\begin{cor}\label{cor2.1}
The Fredholm determinants of $\mathcal C:L^2(\R_+,\C^r)$, $\wh \K:\H^2_r\to \H^2_r$ and $\K:L^2(\gamma_+, \C^p)\to L^2(\gamma_+, \C^p)$ are all equal.
\end{cor}

Given the fact that the operator  $\wh \K $ (with kernel (\ref{khatkern}))  and $\K$ (with kernel (\ref{kernel}))
have the same Fredholm determinant, we shall continue our discussion by focusing on the latter.
In the sequel we will simply analyze operators with kernels as in (\ref{kernel}) and forget about their origin as Fourier transform of convolution operators.
\begin{remark}
	It is worth mentioning that operators with kernel (\ref{kernel}) with $p=r=1$ (and $E_1=E_2$) belong to the same class considered in \cite{TWmKdV}. Using our formalism it is possible to re-derive the connection between the Fredholm determinants of these operators and the mKdV/KdV hierarchies.
\end{remark}

\section{Riemann-Hilbert problems with different asymptotics and their mutual relationship}\label{RHPs}

Given a kernel $\K(\lambda,\mu)$ as in (\ref{kernel}) corresponding to the operator  (denoted by the same symbol)
\be
\K:L^2(\gamma_+,\C^p) \to L^2(\gamma_+,\C^p)\ ,\ \ \ \K(\lambda,\mu):=\frac{ E_1^T(\lambda)E_2(\mu)}{\lambda + \mu} 
\ee
we construct two related Riemann-Hilbert problem on the collection of contours $\gamma:=\gamma_+\cup\gamma_-$ (here $\gamma_-:=-\gamma_+$) and with jump matrix
\be
&\& M(\lambda):= \le[
\begin{array}{cc}
\1_r &-
2i\pi \r(\lambda)\,\chi_{_{\gamma_+}}\\ 
-
2i\pi \wt \r (\lambda)\,\chi_{_{\gamma_-}} &  \1_r
\end{array}
\ri]
\label{3.1}\\[10pt]
&\& \r(\lambda) =
 E_1(\lambda)E_2^T(\lambda)\in \Mat(r\times r)\ ,\ \ \wt \r(\lambda):= \r(-\lambda).
\ee
where $\chi_{_X}$ denotes the indicator function of the set $X$. Here and below we denote with $\sigma_i,i=1,2,3$ the Pauli matrices
\be
\sigma_1:=\le[\begin{array}{cc}0 & 1\\
					1 & 0\end{array}\ri],\quad\quad\sigma_2:=\le[\begin{array}{cc}0 & i\\
					-i & 0\end{array}\ri],\quad\quad
					\sigma_3:=\le[\begin{array}{cc}1 & 0\\
					0 & -1\end{array}\ri],
\nonumber\ee 
$\sigma_+:=(\delta_{i1}\delta_{j2})_{i,j=1,2}$ and $\sigma_-$ its transpose. Furthermore we shall set 
 \be
 \wh \s_k  = \1_r \ot  \s_k\ ,\ \ \ \ k=\pm, 1,2,3,
 \ee
 where by the tensor notation can be taken to mean the matrix of size $2r\times 2r$ split into $2\times 2$ blocks of size $r\times r$.

Note that the  jump matrices $M(\lambda)$ on $\gamma := \gamma_+ \cup \gamma_-$ satisfy
\be
M(-\lambda) = \wh \s_1 M(\lambda) \wh \s_1.\label{32}
\ee
We are going to formulate  two Riemann-Hilbert Problems (Problems  \ref{prob3}, \ref{prob1}) and we will show how they are related between themselves (Prop. \ref{RHPrelation}) and how they relate respectively to the two Fredholm determinants $\det(\Id_{\gamma_+} + \K)$ and $\det(\Id_{\gamma_+}-\K^2)$ and the resolvents of the respective operators. 
In the sequel we shall assume that $E_j(\lambda)$ are  {\bf smooth} (beside the already imposed conditions $E_j \in L^2(\gamma_+ )\cap L^\infty(\gamma_+) \ot  \Mat(r\times p)$).
 
\begin{shaded}
\begin{problem}
\label{prob3}
Find  the sectionally analytic function $\Xi(\lambda)\in GL(2r, \C)$ on $\C\setminus (\gamma_+\cup \gamma_-)$ such that  (with $M(\lambda)$  given in (\ref{3.1}))
\be
\Xi_+(\lambda) &=& \Xi_-(\lambda) M(\lambda)\ \ \ \lambda \in \gamma_+\cup \gamma_-\ \\
\Xi(\lambda) &=& \1_{2r} + \frac {\Xi_1}\lambda + \frac {\Xi_2} {\lambda^2} + \dots  \ ,\ \ \ \lambda\to \infty\ . \label{34}
\ee
\end{problem}
\end{shaded}

\begin{shaded}
\begin{problem}
\label{prob1}
Find  the sectionally analytic function $\Gamma(\lambda)\in GL(2r,\C)$ on $\C\setminus (\gamma_+\cup \gamma_-)$ such that  (with $M(\lambda)$  given in (\ref{3.1}))
\be
\Gamma_+(\lambda) &=& \Gamma_-(\lambda) M(\lambda)\ \ \ \ \lambda \in \gamma_+\cup \gamma_-,\
\label{Gammajumps} \\
\Gamma(\lambda) &=& 
\L (\lambda) \le(\1_{2r} + \sum_{j=1}^{\infty} \frac {\Gamma_j}{\lambda^j}\ri) \ ,\ \ \ \ \lambda\to \infty
\label{expGamma}
\\
\Gamma(\lambda) \L^{-1}(\lambda) &=& \mathcal O(1) \ ,\ \ \ \lambda \to 0
\label{0asy}\\ 
\Gamma(-\lambda) &=& \Gamma(\lambda)\wh \s_1
\label{symmetry2}\\
\Gamma_1 &=& a_1  \ot  \s_3\label{gaugefixing}
\ee 
where the matrix $\L(\lambda)$ is defined as follows
\be
\L(\lambda) := \1_r \ot  L(\lambda) = \le[
\begin{array}{cc}
\1_r & \1_r \\
-i\lambda\1_r & i\lambda\1_r
\end{array}
\ri]\ ,\ \ L(\lambda) := \le[
\begin{array}{cc}
1 & 1 \\
-i\lambda & i\lambda
\end{array}
\ri] \label{elldef}
\ee
\end{problem}
\end{shaded}
The validity of the asymptotic expansions near infinity needs additional conditions on the jump matrices if some component of $\gamma_+$ extends to infinity (if this happens we assume that these components extend to infinity along asymptotic directions) . A sufficient condition, which we hereby tacitly assume, is that $\r(\lambda) = \mathcal O(|\lambda|^{-\infty})$ as $|\lambda|\to\infty$ along any such component and extends to an analytic function on an open sector containing the direction of approach in such a way that the same asymptotic holds. 

It is clear that the two Problems  \ref{prob3}, \ref{prob1} are closely related and the remainder of this section is devoted to explaining their mutual relationship.
It is a straightforward result that, if a solution of Problem \ref{prob3} exists, then it is unique. The uniqueness of the solution for the Problem \ref{prob1} comes from the following proposition.
\begin{prop}
\label{Gammarhp}
Let $\Gamma(\lambda)$ be a sectionally analytic function that solves the  RHP (\ref{Gammajumps}, \ref{expGamma}, \ref{0asy}, \ref{symmetry2}).

If a solution exists then 
\begin{enumerate}
\item $\det \Gamma(\lambda) = (2i\lambda)^r$;
\item Any matrix $\wt \Gamma = (\1_{2r}+ c  \ot  \s_-) \Gamma$ solves the same RHP with $c\in \Mat(r\times r)$, constant;
\item Any solution has an expansion where the terms $\Gamma_j$ in (\ref{expGamma}) have the symmetry
\be
\Gamma(\lambda)  = \L(\lambda) \le(\1_{2r} + \sum_{k=1}^{\infty} \frac {\Gamma_k}{\lambda^k}\ri)\ ,\qquad
\Gamma_j = (-1)^j \wh\s_1 \Gamma_j \wh\s_1
\ee and hence 
\be
\Gamma_{2j} = a_{2j} \ot \1_{2} + b_{2j} \ot  \s_1\ ,\ \ \ \ 
\Gamma_{2j+1} = a_{2j+1}  \ot  \s_3 + b_{2j+1} \ot   \s_2\ ,\ \ \ a_j,b_j\in \Mat(r\times r)
\label{314}
\ee
\item If we additionally require the condition (\ref{gaugefixing})  $\Gamma_1 = a_1 \ot  \s_3$ (for some constant matrix  $a_1$) then the solution is unique. This solution will be referred to as the {\bf gauge-fixed} solution.
\end{enumerate}
\end{prop}
{\bf  Proof of Prop. \ref{Gammarhp}.}
1. It is clear that the determinant  has no jumps because the jump matrices are unimodular. Moreover, from $\det \L=(2i\lambda)^r$ (see (\ref{elldef}))  and (\ref{expGamma}) we have  $\det \Gamma(\lambda) = (2i\lambda)^r(1 + \mathcal O(\lambda^{-1})).$ Finally from (\ref{0asy}) we have $\det\Gamma = \mathcal O(\det \L) =\mathcal O (\lambda^r)$, $\lambda\to 0$, and hence it must be $\det\Gamma \equiv (2i\lambda)^r$.

2. We note that 
\be
(\1_{2r} + c \ot   \s_-) \L(\lambda)= \L(\lambda) \le(\1_{2r} + \frac {ic}{2\lambda} \ot   (\s_3 -i\s_2)\ri) 
\ee
and hence the multiplication on the left by a constant matrix of such a form does not change the form of the asymptotic expansion and does not change the jump conditions. This proves the second point.

3. The statement is obvious once one notices that $\L(-\lambda) =\L(\lambda)\wh \s_1$.

4. Suppose $\wt \Gamma$ is another solution satisfying the same requirements and denote by $\wt a_j, \wt b_j$ the coefficients in its expansion as per (\ref{314}).  By point 1, any two solutions have the same determinant; the ratio
\be
S(\lambda):= \wt \Gamma (\lambda) \Gamma^{-1}(\lambda) 
\ee
must be a holomorphic matrix  function on $\C\setminus \{0\}$.
However, from the condition (\ref{0asy}) we see that actually $S(\lambda)$ must be analytic at $0$ as well. 
 Looking at the behaviour at infinity of $\wt \Gamma$ and $\Gamma$ one finds by a direct computation that $S(\lambda)$ is bounded and 
\be
S(\lambda) = \le[
\begin{array}{cc}
\1_r & 0\\
i a_1 -i\wt a_1 & \1_r
\end{array}
\ri]
\ee

Suppose now that $\wt a_1\neq a_1$; then 
\be
\wt \Gamma(\lambda) = \le[
\begin{array}{cc}
\1_r & 0\\
i a_1 -i\wt a_1 & \1_r
\end{array}
\ri] \Gamma(\lambda)
\ee
But then one sees by direct matrix multiplication that $\wt b_1$ should equal $\frac {\wt a_1-a_1}{2}$ which violates the normalization $\wt b_1=0$. This proves uniqueness.
\QED

In Proposition \ref{RHPrelation} and Proposition \ref{PHRrelation} we  study the relationship between the Riemann--Hilbert problems \ref{prob3} and \ref{prob1}: in particular we shall see that {\bf they are not equivalent}, in the sense that if Problem \ref{prob3} admits a solution then so does  Problem \ref{prob1} but, in general, not viceversa. 
 We start by observing that  the symmetry (\ref{32}) for the jump matrices  implies the same symmetry for $\Xi$
\be
\Xi(-\lambda) = \wh\s_1\Xi(\lambda) \wh\s_1
\ee
which in turns implies the following form for the coefficient $\Xi_j$ in (\ref{34}) 
\be
\Xi(\lambda) = \1_{2r}  + \sum_{k=1}^\infty \frac {\Xi_k}{\lambda^k}\ ,\qquad
\Xi_{2j+1} = \a_{2j+1} \ot  \s_3 + \b_{2j+1} \ot  \s_2\ , \qquad
\Xi_{2j} = \a_{2j} \ot  \1_2 + \b_{2j}  \ot \s_1\ .\label{419}
\ee
\begin{prop}
\label{RHPrelation}
Let $\Xi$ be the solution of Problem \ref{prob3}; then the solution of Problem \ref{prob1} is 
\be
\Gamma(\lambda) = \le[
\begin{array}{cc}
\1_r & \1_r \\
-i\lambda\1_r-2\b_1 & i\lambda\1_r-2\b_1
\end{array}
\ri]\Xi(\lambda)
\ee
with $\b_1$ as in (\ref{419}).
In addition the coefficients of the expansions for $\Gamma$ (\ref{314}) and $\Xi$ (\ref{419}) satisfy
\bea
a_1 = \a_1-i\b_1\ ,\ & \ \ b_1 =0
\label{518}
\\
a_{2j+1} = \a_{2j+1} +  i\b_1 (\b_{2j} - \a_{2j})\  ,&\qquad
b_{2j+1} = \b_{2j+1} +\b_1 \le(\b_{2j} - \a_{2j}\ri)\cr
a_{2j} =  \a_{2j} - i\b_1\le( \a_{2j-1} - i \b_{2j-1} \ri)\ ,&\qquad
b_{2j} = \b_{2j} - i\b_1\le(\a_{2j-1} - i\b_{2j-1}\ri)\label{519}
\eea
\end{prop}
{\bf Proof of Prop. \ref{RHPrelation}.}
Since $\Gamma$ and $\Xi$ have the same jumps we must have 
$
\Gamma(\lambda) = R(\lambda) \Xi(\lambda)
$ for some $R(\lambda)$ at most polynomial.
From the symmetries we must have $R(-\lambda) =R(\lambda)\wh \s_1$ and $\det R = (2i\lambda)^r$. The expansion of $\Gamma$ and $\Xi$ at infinity forces $R$ to be of the form 
\be
R(\lambda) = \le[
\begin{array}{cc} 
\1_r & \1_r \\
-i\lambda\1_r +2ic & i\lambda\1_r+2ic
\end{array}
\ri]
\label{Rmatrix}
\ee
On the other hand, as we presently show, the gauge fixing (\ref{gaugefixing}) determines $C$; indeed
\be
{\mathbf L}^{-1} R =\1_{2r}+ \frac c {\lambda}\ot \le[
\begin{array}{cc}
-1 & -1 \\
1& 1
\end{array}
\ri]  = \1_{2r} + \frac 1 \lambda \le(-c \ot\s_3 + ic\ot \s_2\ri) 
\ee
and therefore  in the expansions of $\Gamma$ and $\Xi$ and matrix multiplications we have
\bea
\1_{2r} +&\&  \sum_{j=0}^\infty \frac {a_{2j+1} \ot\s_3 + b_{2j+1}\ot  \s_2} {\lambda^{2j+1}}+  
\sum_{j=1}^\infty \frac {a_{2j}\ot \1_2 + b_{2j}\ot  \s_1}{\lambda^{2j}}  
=\\
&\&=
\1_{2r} + \frac {(\a_1-c)\ot \s_3 + (\b_1+ic)\ot \s_2}\lambda +\cr
&\&+ \sum_{j=1}^{\infty}
\frac {
\le(\a_{2j} - c (\a_{2j-1} - i   \b_{2j-1})\ri)\ot \1_2 + \le( \b_{2j} - c(\a_{2j-1} - i \b_{2j-1}) \ri)\ot  \s_1
}{\lambda^{2j}}+\nonumber\\
&\&+\sum_{j=1}^{\infty}
\frac{\le( \a_{2j+1} - c \a_{2j} + c \b_{2j} \ri)\ot \s_3 + \le(\b_{2j+1} - i c \b _{2j} + i c \a_{2j}\ri) \ot \s_2
}{\lambda^{2j+1}}
\eea
The gauge fixing (\ref{gaugefixing}) mandates $b_1=0$  (i.e. the coefficient matrix of $\s_2$ in the term $\lambda^{-1}$ must be absent) so that we must have $c = i\b_1$ and equating the coefficients of the expansion above implies (\ref{519}). 
It only remains to show that $R(\lambda)\Xi(\lambda) \L^{-1}(\lambda)$ is bounded at $\lambda=0$ (condition (\ref{0asy})).
Since $\L^{-1} = \1_r \ot L^{-1}$ and $L^{-1}(\lambda)$ has only simple pole at $\lambda=0$, then $R(\lambda)\Xi(\lambda) \L^{-1}(\lambda) = \frac {c_0}\lambda + \mathcal O(1)$ as $\lambda\to 0$. On the other hand the symmetries imply $R(\lambda)\Xi(\lambda) \L^{-1}(\lambda) =R(-\lambda)\Xi(-\lambda) \L^{-1}(-\lambda)$ and hence $c_0=-c_0$ so that $c_0=0$. 
\QED
\begin{prop}
\label{PHRrelation}
Let $\Gamma(\lambda)\in GL(2r,\C)$ be the solution of Problem \ref{prob1} and denote the $r\times r$ blocks of $\Gamma$ by $\Gamma_{ij}$, $i,j=1,2$; then the solution of Problem \ref{prob3} for $\Xi$  exists if and only if 
$$
\det \Gamma_{11} (0) \neq  0\ .
$$
Moreover 
$$
\b_1 = -\frac 1 2{\Gamma_{11}^{-1}(0)} {\Gamma_{21}(0)}=-i \lim_{\lambda\to\infty}\lambda \Xi_{12}(\lambda)
$$
\end{prop}
{\bf Proof of Prop. \ref{PHRrelation}.}
Let $\Gamma(\lambda)$ be the solution of Problem \ref{prob1}. In particular $\Gamma(\lambda)$ is bounded everywhere (by definition) and we want now to find a matrix $R(\lambda)$ of the form (\ref{Rmatrix}) such that 
\be
\Xi(\lambda):= R^{-1}(\lambda)\Gamma(\lambda)
\ee
solves Problem \ref{prob3}. It is clear that the jumps will be automatically satisfied and so the asymptotic behaviour at infinity. The value of the constant matrix  $c$ must be determined by the requirement that $\Xi$ is {\em bounded} at $\lambda=0$.
From the symmetry (\ref{symmetry2}) we have the matrix equations
\be
\Gamma_{11}(0) = \Gamma_{12}(0)\ ,\ \ \Gamma_{21}(0) = \Gamma_{22}(0)\ .
\label{430}
\ee
A direct computation yields
\bea
R^{-1}(\lambda) \Gamma(\lambda) = \mathcal O(1) \ ,\ \ \ \lambda\to 0\\
\le[
\begin{array}{cc}
\frac {\1_r} 2  + \frac {c}\lambda  &  -\frac {\1_r}{2i\lambda}\\
\\
\frac {\1_r} 2 - \frac {c}\lambda  & \frac {\1_r}{2i\lambda}
\end{array}
\ri] \Gamma(\lambda) =\mathcal O(1)\\
c \Gamma_{11}(0) +\frac i 2 \Gamma_{21}(0) =0\ ,\Rightarrow \ \ 
c = -\frac i 2 \Gamma_{11}^{-1}(0)  \Gamma_{21}(0)\label{330}\\
c \Gamma_{12}(0) +\frac i 2 \Gamma_{22}(0) =0\ ,
\eea
The two equations are the same due to (\ref{430}).
Now, if $\det \Gamma_{11}(0)\neq 0$ then the solution for $c$ is as in (\ref{330}) and the sufficiency is proved.
As for the necessity, if $\det \Gamma_{11}(0)=0$, then the equation (\ref{330}) may still be compatible. However this would mean that there are infinitely many $c$ that solve the matrix equation (\ref{330}), which would violate the uniqueness of  the RHP \ref{prob3}. \QED

We conclude the section with the following two theorems, which we state side-by-side for the sake of easy comparison.
\begin{shaded}
\begin{theorem}
\label{thmXi}
Let $\K(\lambda,\mu)$ be the integral operator on $L^2(\gamma_+,\C^p)$ with kernel 
\be
\mathcal K(\lambda,\mu):= \frac{E_1(\lambda)^TE_2(\mu)}{\lambda+\mu}\label{Kzwt}
\ee 
Then the resolvent operator $\mathcal R_{++} = \K^2\circ(\Id_{\gamma_+} -\K^2)^{-1}$ of $\Id_{\gamma_+} - \K^2$ on $L^2(\gamma_+,\C^p)$  has kernel $\mathcal R_{++}(\lambda,\mu)$ given by 
\be
\mathcal R_{++}(\lambda,\mu) =  
[
E_1^T(\lambda)
,
\0_{p\times r}
] 
\frac{\Xi^{T}(\lambda)\Xi^{-T}(\mu)}{\lambda - \mu}\le[
\begin{array}{c}
\0_{r\times p}
\\
\!\!\! E_2(\mu)\!\!\! 
\end{array}
\ri]
\ee
where $\Xi$ is the solution of Problem \ref{prob3}  with the jump matrix (\ref{3.1}).

If  $\r(\lambda):= E_1(\lambda)E_2^T(\lambda)$ is symmetric, $\r=\r^T$ (for example if $E_1=E_2=E$) 
then the resolvent can be written more symmetrically as 
\be
\mathcal R_{++}(\lambda,\mu) = 
i [
E_1^T(\lambda)
,
\0_{p\times r}
] 
\frac{\Xi^T(\lambda)\wh \s_2 \Xi(\mu)}{\lambda - \mu}\le[
\begin{array}{c}
\!\!\! E_2(\mu)\!\!\! 
\\
\0_{r\times p}
\end{array}
\ri]
\ee
The solution to Problem \ref{prob3} exists if and only if the operator $\Id_{\gamma_+} -\K^2$ is invertible.
\end{theorem}
\end{shaded}
\begin{shaded}
\begin{theorem}
\label{thmGamma}
Let $\K(\lambda,\mu)$ be the integral same operator as in Thm. \ref{thmXi}. 
Then the resolvent operator $\mathcal S = -\K\circ (\Id_{\gamma_+} + \K)^{-1}$ has kernel $\mathcal S(\lambda,\mu)$ given by 
\be
S(\lambda,\mu)  =
\frac {2 \mu \le[E_1^T(\lambda),\0_{p\times r} \ri] \Gamma^{T}(\lambda) \Gamma^{-T}(\mu) 
\le[\begin{array}{c}
\0_{r\times p} 
\\
E_2(\mu)
\end{array} \ri]}{\lambda^2  - \mu^2}
\ee
where $\Gamma$ is the solution of Problem \ref{prob1} with the jump matrix (\ref{3.1}).
If  $\r(\lambda):= E_1(\lambda)E_2^T(\lambda)$ is symmetric, $\r=\r^T$ (for example if $E_1=E_2=E$) then the resolvent can be written more symmetrically as 
\be
\mathcal S(\lambda,\mu) =
\frac{ \le[
 E_1^T(\lambda)
  ,
 \0_{p\times r}
   \ri] \Gamma^{T}(\lambda) \wh \s_2 \Gamma (\mu) 
\le[\begin{array}{c}
  E_2(\mu)
\\
\0_{r\times p} 
\end{array} \ri]}{\lambda^2  - \mu^2}
\ee

 This solution to Problem \ref{prob1} exists if and only if the operator $\Id_{\gamma_+} +
 \K$ is invertible.
\end{theorem}
\end{shaded}
{\bf Proof of Thm. \ref{thmXi}.}
We start observing that the jump $M(\lambda)$ in Problem \ref{prob3} can be written as
\bea
M(\lambda) &\&= \1 
-
 2i\pi {\bf f}(\lambda) {\bf g}^{\mathrm{T}}(\lambda)\\
{\bf f}(\lambda) &\&= \le[
\begin{array}{c}
E_1(\lambda)\\ \0_{r\times p}
\end{array}
\ri]\chi_{_{\gamma_+}}(\lambda) + \le[
\begin{array}{c}
\0_{r\times p} \\ \wt E_1 (\lambda)
\end{array}
\ri]\chi_{_{\gamma_-}}(\lambda) \\
{\bf g}(\lambda) &\&= \le[
\begin{array}{c}
\0_{r\times p}\\ 
E_2(\lambda)
\end{array}
\ri]\chi_{_{\gamma_+}}(\lambda) + \le[
\begin{array}{c} 
\wt E_2(\lambda)\\ \0_{r\times p} 
\end{array}
\ri]\chi_{_{\gamma_-}}(\lambda)\ ,\ \ \ \ \wt E_j(\lambda):= E_j(-\lambda).
\eea
By the IIKS theory, this RHP is associated to the kernel $N$ acting on $L^2(\gamma_+\cup\gamma_-, \C^p)$ with kernel given by
\be
	N(\lambda,\mu)=
	\frac{{\bf f}^{\mathrm T}(\lambda){\bf g}(\mu)}{\lambda-\mu}=
\frac{E_1^T(\lambda)\wt E_2(\mu) \chi_{_{\gamma_+}}(\lambda)\chi_{_{\gamma_-}}(\mu)+ \wt E_1^T(\lambda)E_2(\mu)  \chi_{_{\gamma_-}}(\lambda)\chi_{_{\gamma_+}}(\mu)}{\lambda-\mu}\label{361}
\ee 
According to the split $L^2(\gamma_+\cup\gamma_-)= L^2(\gamma_+)\oplus L^2(\gamma_-)$, using the naturally related matrix notation, 
we can write $N$ as 
\be
N =  \le[\begin{array}{c|c}
0 & \mathcal G\\
 \hline
\mathcal F & 0
\end{array}\ri]\ ,\ \ \mathcal G:L^2(\gamma_-,\C^p)\to L^2(\gamma_+,\C^p) \ ,\ \ \ \mathcal F:L^2(\gamma_+,\C^p)\to L^2(\gamma_-,\C^p) 
\ee
where the operators $\mathcal F$ and $\mathcal G$ are integral operators with kernels
\bea
	\mathcal G(\lambda,\xi)=
	 \frac{ E_1^T(\lambda ) \wt E_2(\xi)\chi_{_{\gamma_+}}(\lambda)\chi_{_{\gamma_-}}(\xi)}{\lambda-\xi}      \ ,
	\qquad 
	\mathcal F(\xi, \mu)=
	\frac{\wt E_1^T (\xi) E_2(\mu) \chi_{_{\gamma_-}}(\xi)\chi_{_{\gamma_+}}(\mu)}{\xi-\mu}\label{FG}
\eea
We observe that the kernel of the composition reads
\bea
(\mathcal G \circ \mathcal F)(\lambda,\mu)=
E_1^T(\lambda)\le(\int_{\gamma_-}\frac{\wt E_2(\xi)\wt E_1^T(\xi)}{(\lambda-\xi)(\xi-\mu)}d\xi\ri) E_2(\mu)=\\
E_1^T(\lambda)
\le(\int_{\gamma_+}\frac{E_2 (\xi)E_1^T (\xi)\d\xi}{(\lambda+\xi)(\xi+\mu)}\ri)  
E_2(\mu)
= \mathcal K^2(\lambda,\mu)
	 \label{composition}
\eea
and hence our task of computing the resolvent of $\Id_{\gamma_+} - \mathcal K^2$ is the same as computing the resolvent of $\Id_{\gamma_+} - \mathcal G\circ \mathcal F$.
To this end we write first the resolvent of $\Id_{\gamma_+\cup \gamma_-} - N$ (using \cite{IIKS}):
\be
\mathcal R(\lambda,\mu) = \frac {{\bf f}^T(\lambda) \Xi^{T}(\lambda) \Xi^{-T}(\mu) {\bf g}(\mu)}{\lambda - \mu} 
\ee
according to the projections in $L^2(\gamma_\pm, \C^p)$:
\bea
\mathcal R(\lambda,\mu) &\& =
\mathcal R_{++}(\lambda,\mu)
+
\mathcal R_{-+}(\lambda,\mu)
+
\mathcal R_{+-} (\lambda,\mu)
+
\mathcal R_{--}(\lambda,\mu)\nonumber\\
=
 [E_1^T(\lambda),&\&\0 _{p\times r}] 
\frac{\Xi^{T}(\lambda)\Xi^{-T}(\mu)}{\lambda - \mu}\le[
\begin{array}{c}
 \0_{r\times p}
 \\
 \!\!\!
  E_2(\mu)\!\!\!
\end{array}
\ri]\!\! \chi_{_{\gamma_+}}\!\!(\lambda)\chi_{_{\gamma_+}}\!\!(\mu)
+  
[\0_{p\times r},\! \wt E_1^T(\lambda)] \frac{\Xi^{T}(\lambda)\Xi^{-T}(\mu)}{\lambda - \mu}\!\! \le[
\begin{array}{c}
\0_{r\times p}\\\!\!\! 
E_2(\mu)\!\!\!
\end{array}
\ri]\!\! \chi_{_{\gamma_-}}\!\!(\lambda)\chi_{_{\gamma_+}}\!\! (\mu)
+
\nonumber
\\
+
 [ E_1^T(\lambda),&\&\0_{p\times r} ]
  \frac{\Xi^{T}(\lambda)\Xi^{-T}(\mu)}{\lambda - \mu}\le[
\begin{array}{c}
\!\!\!
\wt E_2(\mu)\!\!\!\\ \0_{r\times p}
\end{array}
\ri]\chi_{_{\gamma_+}}\!\! (\lambda)\chi_{_{\gamma_-}}\!\! (\mu) 
+
[    \0_{p\times r} , \wt E_1^T(\lambda) ]
\frac{\Xi^{T}(\lambda)\Xi^{-T}(\mu)}{\lambda - \mu}\le[
\begin{array}{c}
\!\!\! 
\wt E_2(\mu)\!\!\!
\\
\0_{r\times p}
\end{array}
\ri]\chi_{_{\gamma_-}}\!\!(\lambda)\chi_{_{\gamma_-}}\!\!(\mu) \nonumber\\
\label{final}
\eea
where the four addenda appears in the matrix notation induced by the splitting $L^2(\gamma_+\cup\gamma_-)=L^2(\gamma_+)\oplus L^2(\gamma_-)$;
$$(\Id_{\gamma_+\cup\gamma_-}-K)^{-1}=\le[\begin{array}{ccc}
									\Id_{\gamma_+}+\mathcal R_{++} & \mathcal R_{+-}\\
									\mathcal R_{-+} & \Id_{\gamma_-} + \mathcal R_{--}
									\end{array}\ri] $$
On the other hand we have
\bea
	\le[\begin{array}{ccc}
		(\Id_{\gamma_+}-\mathcal G \circ \mathcal F)^{-1} & 0\\
		\mathcal F\circ (\Id_{\gamma_+}-\mathcal G \circ \mathcal F)^{-1} & \Id_{\gamma_-}\end{array}\ri]=\le[\begin{array}{ccc}
\Id_{\gamma_+}-\mathcal G \circ \mathcal F & 0\\
-\mathcal F & \Id_{\gamma_-} \end{array}\ri]^{-1}=
\nonumber\\
=\le[\begin{array}{ccc} \Id_{\gamma_+} & -\mathcal G\\
-\mathcal F & \Id_{\gamma_-}\end{array}\ri]^{-1}\circ\le[\begin{array}{ccc} \Id_{\gamma_+} & -\mathcal G\\
0 & \Id_{\gamma_-}\end{array}\ri]=
\le[\begin{array}{ccc} \Id_{\gamma_+}+\mathcal R_{++} & \mathcal R_{+-}\\
\mathcal R_{-+} & \Id_{\gamma_-}+\mathcal R_{--}\end{array}\ri]\circ\le[\begin{array}{ccc} \Id_{\gamma_+} & -\mathcal G\\
0 & \Id_{\gamma_-}\end{array}\ri]\nonumber
\eea
so that the entry $(1,1)$ of the equation above gives
$$(\Id_{\gamma_+}-\mathcal G \circ \mathcal F)^{-1}=\Id_{\gamma_+}+\mathcal R_{++}$$ and the equation (\ref{final}) gives the precise form of the kernel $\mathcal R_{++}(\lambda,\mu)$.

In case of symmetry $\r=\r^T$ this form simplifies because $\Xi^{-1}(\lambda) = \wh \s_2 \Xi^T(\lambda) \wh \s_2$ (which is proved along the same lines as in Thm. \ref{thmGamma}). The statement about the existence is a direct application of IIKS theory.
 \QED

{\bf Proof of Thm. \ref{thmGamma}.}
The idea of the proof is to reduce as much as possible the theorem to the theory of integrable operators of Its-Izergin-Korepin-Slavnov (IIKS). We can write the operator as 

\be
\mathcal K(\lambda,\mu):= \frac{E_1(\lambda )^TE_2(\mu)}{\lambda+\mu}=  \frac{(\lambda-\mu) E_1(\lambda)^TE_2(\mu)}{\lambda^2-\mu^2}\label{Kzw}
\ee
We now introduce the coordinate $z:= \lambda^2$ and $w:=\mu^2$. Since  $\gamma_+$ is in the upper half-plane, its image under the square map is well-defined and lies in $\C\setminus \R_+$.
Since the arc-length of $\gamma_+$ differs in the $z$-plane and $\lambda$-plane, we must introduce the square-roots of the Jacobians. The integral operator (\ref{Kzw}) reads 
\be
\mathcal K(z,w ):= 
\frac{(\sqrt{z} - \sqrt w) E_1(\sqrt{z} )^TE_2(\sqrt{w} )} { 2 (zw )^\frac 1 4 (z - w)}
= 
\frac{-\le[ E_1^T(\sqrt{z}) , -i\sqrt{z} E_1^T(\sqrt{z}) \ri]
 \le[
\begin{array}{c}
E_2(\sqrt{w})\\
-
\frac i{\sqrt {w}} E_2(\sqrt{w})
\end{array}
\ri]
 } { 2  (z - w)} \le(\frac w z\ri)^\frac 14
\ee
We have to construct the resolvent of $\Id_{\gamma_+} + \K = \Id_{\gamma_+} - (-\K)$. 
We have now an integrable kernel in the sense of Its-Izergin-Korepin-Slavnov where 
the matrices $\bf f, \bf g$ can be chosen as   
\be
(-\K)(z,w) = \frac {{\bf f}^T(z) {\bf g}(w)}{z-w}\ ,\ \  
{\bf f}(z) =\frac1 {\sqrt[4]{z}} \le[
\begin{array}{c}
E_1(\sqrt{z}) \\
-i\sqrt{z}E_1(\sqrt{z})
\end{array} \ri]\ ,\ \ \ \ 
{\bf g}(z) = \frac 1 2 \sqrt[4]{z} \le[
\begin{array}{c}
E_2(\sqrt{z})\\
-
\frac{i  E_2(\sqrt{z})}{\sqrt{z}}
\end{array}
\ri]
\ee 
We immediately observe that 
\be
{\bf f}(z) =\frac 1{\sqrt[4]{z}}  \L(\sqrt{z}) \le[
\begin{array}{c}
E_1(\sqrt{z})
\\\0_{r\times p}
\end{array}
\ri]=: \L(\sqrt{z}) {\bf f}_0(z)\ ,\ \ \ \ \\
{\bf g}(z) =\sqrt[4]{z} ( \L^{-1})^T(\sqrt{z}) \le[
\begin{array}{c}
\0_{r\times p}\\
E_2(\sqrt{z}) 
\end{array}
\ri]
=
\L^{-T}(z) {\bf g}_0(z).
\ee
The construction of the resolvent is then associated, in the standard way \cite{IIKS}, to the following Riemann--Hilbert Problem
\bea
\Theta(z)_+ =\Theta(z)_- \le(\1_{2r}  
-
 2i\pi
{\bf f}(z) {\bf g}^T(z) \ri) \ ,\ \ \ z\in \gamma_+\\
\Theta(z) = \1_{2r} + \mathcal O(z^{-1}) \ ,\ \ \ z \to \infty
\eea

We can rewrite the jump matrix as follows 
\be
\1_{2r}  - 2i\pi \L(\sqrt{z} )  (\r(\sqrt z) \ot  \s_+) \L^{-1}(\sqrt{z}) \\
\r(\sqrt{z}):= 
E_1(\sqrt{z}) E_2^T(\sqrt{z})  
\ee
Consequently we introduce the new matrix $\Theta(z ) \L(\sqrt{z} )$, where $\L(\sqrt{z}):= \1_r \ot  L(\sqrt z)$. 
In order to connect with Problem \ref{prob1} we define 
\be
\wh \Gamma(\lambda) = 
\Theta(\lambda^2 ) \L(\lambda) 
\ee
and we see immediately that $\wh \Gamma(-\lambda) = \wh \Gamma(\lambda) \wh \s_1$. Furthermore $\wh \Gamma(\lambda) \L^{-1}(\lambda) =\Theta(\lambda^2)= \mathcal O(1)$ as $\lambda\to 0$.
Thus $\wh \Gamma$ solves Problem \ref{prob1} except for the gauge-fixing (\ref{gaugefixing}), which we now take into consideration:
if we denote by $a_1$ the $(1,2)$ block of size $r\times r$ in 
\be
\Theta(z) = \1 + \frac 1 z \le[
\begin{array}{c|c}
\star &- a_1\\
\hline 
\star & \star
\end{array}
\ri] + \mathcal O(z^{-2})
\ee
then one verifies by matrix multiplication that the relation between $\wh \Gamma$ and the gauge--fixed $\Gamma$ is 
\be
\le(
\1_{2r} + a_1 \ot  \s_+
\ri)\wh \Gamma(z) =  \Gamma(z)\ ,
\ee

The resolvent operator, according to the general theory, is 
\be
\mathcal S(\lambda,\mu) = \frac{{\bf f}^T(z) \Theta^{T}(z) \Theta^{-T}(w){\bf g }(w)}{z-w } \sqrt{\d z\d w} = \\
=\frac {{\bf f}_0^T(z) \L^T(\lambda)\Theta^{T}(\lambda^2) \Theta^{-T}(\mu^2)\L^{-T}(\mu) {\bf g_0}(w)}{z-w} \sqrt{\d z\d w} =\\
=2\sqrt{\lambda\mu}\frac {{\bf f}_0^T(\lambda) \Gamma^{T}(\lambda) \Gamma^{-T}(\mu) {\bf g }_0(\mu)}{\lambda^2-\mu^2} \sqrt{\d \lambda\d \mu} =\\
=
\frac {2 \mu \le[
 E_1^T(\lambda) 
,
\0_{p\times r}
 \ri] \Gamma^{T}(\lambda) \Gamma^{-T}(\mu) 
\le[\begin{array}{c}
\0_{r\times p} 
\\
E_2(\mu)
\end{array} \ri]}{\lambda^2  - \mu^2}
\ee
Finally, note that
\be
\L^{-T}(\lambda) =\frac 1{2i\lambda}\wh \s_2  \L(\lambda ) \wh \s_2 
\ee
and if  $\r=\r^T$ then 
\be
\Gamma^{-T} (\mu) =\frac 1{2i\mu}\wh \s_2 \Gamma (\mu) \wh \s_2
\ee
which can be checked by verifying that ${2i\mu}\wh \s_2 \Gamma^{-T} (\mu) \wh \s_2$ solves the same Problem \ref{prob1} and hence equals $\Gamma$.
In this case the formula for the resolvent takes a more symmetric form 
\be
\mathcal S(\lambda,\mu) =
\frac {
 \le[
  E_1^T(\lambda) 
 , 
 \0_{p\times r}
 \ri] \Gamma^{T}(\lambda) \wh \s_2 \Gamma (\mu) 
\le[\begin{array}{c}
  E_2(\mu)
\\
\0_{r\times p} 
\end{array} \ri]}{\lambda^2  - \mu^2}
\ee

As for the statement of existence;  $\Gamma$ exists if and only if $\Theta$ exists, which is equivalent to the invertibility of the mentioned operator by the IIKS general theory.
\QED

\section{Tau functions and Fredholm determinants}\label{taufunctions}

Slightly generalizing the definition in \cite{BertolaIsoTau} (which is itself a generalization of the notion of isomonodromic tau function introduced in the work of Jimbo-Miwa-Ueno \cite{JMU1,JMU2,JMU3}) we associate, to the space of deformations of the  Riemann--Hilbert problems \ref{prob3} and \ref{prob1} the two differentials below. 
\begin{shaded}
\begin{defn}
\label{defforms}
We define the two forms over the space of deformations of Problem \ref{prob1} and  Problem \ref{prob3}
\be
\omega_{_\Xi}(\pa):= \int \Tr \le(\Xi_-^{-1}\Xi'_- \pa M M^{-1}\ri) \frac{ \d\lambda}{2i\pi}
\label{omegaxi}
\ee
and
\be
\omega_{_\Gamma}(\pa):= \frac 12 \int \Tr \le(\Gamma_-^{-1}\Gamma'_- \pa M M^{-1}\ri) \frac{ \d\lambda}{2i\pi}
\label{omegagamma}
\ee
where $\pa$ denotes any deformation of the jump matrices, ' the derivatives with respect to the spectral parameter and the integration is extended to all the contours where the jumps are supported, $\gamma_+\cup \gamma_-$.
\end{defn}
\end{shaded}
In the cases in which these two differential forms are closed it is defined, up to a constant, the corresponding tau function given by
$\pa\ln\tau_{\Xi/\Gamma}=\omega_{\Xi/\Gamma}(\pa).$

A particular case (of great interest) of deformations is when the jump matrices have the form  
\be
M(\lambda;s) = {\rm e}^{T(\lambda)} M_0(\lambda) {\rm e}^{-T(\lambda)}\ ,
\label{sdep}
\ee
and $T(\lambda)$ is a {\bf diagonal} matrix depending on deformation parameters, while $M_0(\lambda)$ is assumed independent of them. A typical case is $T(\lambda) = \sum_{k=0}^N T_k\lambda^k$ and the diagonal matrices $T_k$ are taken as deformation parameters.

The relation between the Definition \ref{defforms} and Fredholm determinants is elucidated in the following two theorems, stated side-by-side for comparison.
\begin{shaded}
\begin{theorem}\label{4.1}
Given an operator $\K$ as in Section \ref{convolution} and the Riemann-Hilbert Problem \ref{prob3} (with the same $\r(\mu)$) we have the equality
\be
	\pa \ln\tau_{_\Xi} = \int_{\gamma_+\cup \gamma_-} \Tr \le( \Xi_-^{-1} \Xi_-' \pa M M^{-1}
	\ri) \frac {\d\lambda}{2i\pi}= \pa \ln \det (\Id_{\gamma_+} - \mathcal K^2)
\ee
\end{theorem}
\end{shaded}

\begin{shaded}
\begin{theorem}
\label{4.2}
Given an operator $\K$ as in Section \ref{convolution} and the Riemann-Hilbert Problem  \ref{prob1} (with the same $\r(\mu)$) we have the equality
\be
	\pa \ln\tau_{_\Gamma} =\frac 1 2 \int_{\gamma_+\cup \gamma_-} \Tr \le( \Gamma_-^{-1} \Gamma_-' \pa M M^{-1}
	\ri) \frac {\d\lambda}{2i\pi}= \pa \ln \det (\Id_{\gamma_+}
	 +
	 \mathcal K)
\ee
\end{theorem}
\end{shaded}

{\bf Proof of Thm. \ref{4.1}.}
In \cite{BertolaCafasso1} (see Theorem 2.1) it was proved (for the case of scalar operators, but the proof does not differ significantly as we see below) that 
\be\label{previousarticle}
	\pa\ln\tau_{\Xi}=\pa\ln\det(\Id_{\gamma_+\cup\gamma_-}-N)
\ee 
where the integral operator $N$, acting on $L^2(\gamma_+\cup\gamma_-)$
is the one expressed in (\ref{361}). 
On the other hand we have the identity 
\be
\det (\Id_{\gamma_+\cup\gamma_-}-N) = \det (\Id_{\gamma_+}-\mathcal F\circ \mathcal G ) =\det (\Id_{\gamma_+}-\K^2 )
\ee
where the first equality follows from 
\be
\det\le(\Id_{\gamma_+\cup\gamma_-} - \le[
\begin{array}{cc}
0& \mathcal G\\
\mathcal F & 0
\end{array}
\ri]\ri) = \det\le(\Id_{\gamma_+\cup\gamma_-} - \le[
\begin{array}{cc}
0& \mathcal G\\
\mathcal F & 0
\end{array}
\ri]\ri) \det  \le[
\begin{array}{cc}
\Id_{\gamma_-} & \mathcal G\\
0 & \Id_{\gamma_+}
\end{array}
\ri]=\\
 \det\le(\Id_{\gamma_+\cup\gamma_-} - \le[
\begin{array}{cc}
0&0\\
\mathcal F & \mathcal F\circ \mathcal G
\end{array}
\ri]\ri)  = \det \le(\Id_{\gamma_+} - \mathcal F\circ \mathcal G\ri).
\ee
and the second is (\ref{composition}). 
The above computation is formal inasmuch as one would need to prove that all the operators involved are of trace-class. To see that we now prove that both $\mathcal F, \mathcal G$ are trace-class in $L^2(\gamma_+\cup\gamma_-,\C^p)$. Recalling their definition (\ref{FG}) we augment the Hilbert space as $\wh {\mathcal H}:= L^2(\gamma_+\cup\gamma_-,\C^p) \oplus L^2(\R, \C^r)$, and extend trivially  the definition of  $\mathcal F, \mathcal G$ to the augmented space. This allows to represent them as the composition of two Hilbert--Schmidt operators (thus immediately implying the trace class property). Indeed (for example for $\mathcal F$) we have the identity below
\be
\mathcal F(\xi, \mu)=
	\frac{\wt E_1^T (\xi) E_2(\mu) \chi_{_{\gamma_-}}(\xi)\chi_{_{\gamma_+}}(\mu)}{\xi-\mu} = 
\int_\R \frac{\d \zeta}{2i\pi} \frac{\wt E_1^T (\xi)\chi_{_{\gamma_-}}(\xi)}{(\xi-\zeta)} \frac {E_2(\mu) \chi_{_{\gamma_+}}(\mu)}{(\zeta-\mu)} 
\ee
which follows from Cauchy's residue theorem by closing the $\zeta$ integration either in the upper or in the lower half-plane. This realizes $\mathcal F$ as the composition of two operators between the subspaces $L^2(\gamma_+,\C^p)\to L^2(\R, \C^r) \to L^2(\gamma_-, \C^p)$, each of which is Hilbert--Schmidt:
\be
\int_{\gamma_-}|\d\xi| \int_\R |\d\zeta| \frac {\Tr \le(\wt E_1^\dagger(\xi)\wt E_1(\xi)\ri)}{|\xi-\zeta|^2} <\infty>  \int_{\gamma_+}|\d\mu| \int_\R |\d\zeta| \frac {\Tr \le(E_2^\dagger(\mu)E_2(\mu)\ri)}{|\mu-\zeta|^2} 
\ee

For the sake of self-containedness we shall re-derive (\ref{previousarticle}) below.   
let us denote by $\Xi =[\A,\B]$ the two block-columns of $\Xi$ (of sizes $2r\times r$) and by $\Xi^{-1} = \le[
\begin{array}{c}
{\bf C}\\
{\bf D}
\end{array}
\ri]$ the block rows of $\Xi^{-1}$ (of sizes $r\times 2r$). Then
\bea
 \int_{\gamma_+\cup \gamma_-} \Tr \le( \Xi_-^{-1} \Xi_-' \pa M M^{-1}
	\ri) \frac {\d\lambda}{2i\pi} =\\-
 \int_{\gamma_+} \Tr \le( \le[
\begin{array}{c}
{\bf C}_-\\
{\bf D}_-
\end{array}
\ri][\A'_-,\B'_-] \pa  \r \ot \s_+\ri)\d\lambda -
 \int_{\gamma_-} \Tr \le( \le[
\begin{array}{c}
{\bf C}_-\\
{\bf D}_-
\end{array}
\ri][\A'_-,\B'_-] \pa  \wt \r \ot \s_-\ri)\d\lambda
=\\-
 \int_{\gamma_+} \Tr \le(
{\bf D}\A'  \pa \r \ri)\d\lambda 
-
 \int_{\gamma_-} \Tr \le(
{\bf C}\B'  \pa \wt \r \ri)\d\lambda 
\eea
where we have used that ${\bf C}$ and $\A$ are analytic across $\gamma_+$ and ${\bf D}, \B$ analytic across $\gamma_-$.  
On the other hand the jump relations imply
\be
\A(\lambda) = \le[\begin{array}{c}
\1_r\\
\0_{r\times r}
\end{array}\ri] -
\int_{\gamma_-} \frac {\B(\mu) \wt \r(\mu)\d\mu}{\mu-\lambda}\ ,\ \ 
 \B(\lambda) = \le[\begin{array}{c}
\0_{r\times r}\\
\1_r
\end{array}\ri] -
\int_{\gamma_+} \frac {\A(\mu) \r(\mu)\d\mu}{\mu-\lambda}
\ee
and these identities can be differentiated on $\gamma_+$ (for $\A$) and $\gamma_-$ (for $\B$). We thus have 
\be
-
 \int_{\gamma_+} \Tr \le(
{\bf D}\A'  \pa \r\ri)\d\lambda -
 \int_{\gamma_-} \Tr \le(
{\bf C}\B'  \pa \wt \r \ri)\d\lambda 
=\\
=
 \int_{\gamma_+}\!\!\!\! \d\lambda \!\!\! \int_{\gamma_-} \!\!\!\!\! \d\mu \Tr \le(
{\bf D}(\lambda) \frac {\B(\mu) \wt \r(\mu)}{(\mu-\lambda)^2}   \pa \r(\lambda)\ri)\d\lambda
+
 \int_{\gamma_-}\!\!\!\! \d\lambda \!\!\! \int_{\gamma_+} \!\!\!\!\! \d\mu \Tr \le(
{\bf C}(\lambda) \frac {\A(\mu) \r(\mu)}{(\mu-\lambda)^2}   \pa \wt \r(\lambda)\ri) 
\ee
On the other hand these two terms exactly compute \footnote{here we used the symmetries 
$$\B(\lambda)=\wh \sigma_1\A(-\lambda),\quad\quad\quad \D(\lambda)={\bf C}(-\lambda)\wh \sigma_1.$$}
\be
- {\Tr}_{\gamma_+} \le(\mathcal R_{+-}\circ \pa \mathcal F\ri) - {\Tr}_{\gamma_-} \le(\mathcal R_{-+}\circ \pa \mathcal G\ri) = -{\Tr}_{\gamma_+\cup\gamma_-} \le(\le(\Id_{\gamma_+\cup\gamma_-} + \mathcal R\ri)\circ \pa N\ri) =\nonumber\\
= \pa\ln \det(\Id_{\gamma_+\cup \gamma_-} - N)
\ee
where $\mathcal R$ is the resolvent of the operator with kernel $N$ already used in (\ref{361}) and $\mathcal R$ has been decomposed as in (\ref{final}) and $\mathcal F, \mathcal G$ defined in (\ref{FG}). \QED
{\bf Proof of Thm. \ref{4.2}}
We start analyzing the l.h.s. of the equation and observing that $\pa\ln\tau_\Gamma$ is equal to $\frac{1}2\pa\ln\tau_\Xi$ plus an additional term. Indeed, from $ \Gamma = R(\lambda) \Xi(\lambda)$ with $R$ defined in (\ref{Rmatrix}),  we find 
\be
\Gamma^{-1} \Gamma' = \Xi^{-1} \Xi'  + \Xi^{-1} R^{-1} R' \Xi = \Xi^{-1} \Xi'   + \frac 1{2\lambda} \Xi^{-1}(\lambda ) (\1 - \wh\s_1) \Xi(\lambda )
\ee
so that 
\bea
2\pa\ln\tau_\Gamma-\pa\ln\tau_\Xi=\int_{\gamma_+  \cup \gamma_-} \Tr \le(\Gamma_{-}^{-1} \Gamma_-' \pa M M^{-1}\ri) \frac {\d \lambda}{2i\pi} 
-
 \int_{\gamma_+  \cup \gamma_-} \Tr \le(\Xi_{-}^{-1} \Xi_-' \pa M M^{-1}\ri) \frac {\d \lambda}{2i\pi} 
=\nonumber\\
 =
 \int_{\gamma_+  \cup \gamma_-} \frac 1{2\lambda}\Tr \le(\Xi_{-}^{-1}(\lambda)\wh  \s_1 \Xi_-(\lambda) \le[
\begin{array}{cc}
\0_{r\times r} & \pa \r \,\chi_+\\
\pa \wt \r\, \chi_- & \0_{r \times r}
\end{array}
\ri]\ri)  \d \lambda 
=\nonumber\\
= 
\int_{\gamma_+} \frac 1{2\lambda}\Tr \le(\Xi_{-}^{-1}(\lambda) \wh \s_1 \Xi_-(\lambda) \wh \s_+\pa \r(\lambda)\ri)  \d \lambda  +
\int_{\gamma_-} \frac 1{2\lambda}\Tr \le(\Xi_{-}^{-1}(\lambda) \wh \s_1 \Xi_-(\lambda) \wh \s_-\pa \wt \r(\lambda) \ri)\d \lambda=\nonumber\\
=
\int_{\gamma_+} \frac 1{\lambda}\Tr \le(\Xi_{-}^{-1}(\lambda) \wh \s_1 \Xi_-(\lambda) \wh \s_+ \pa \r(\lambda) \ri)\d \lambda =
 \int_{\gamma_+} \frac 1{\lambda}\Tr \le(\mathbf D(\lambda) \wh \s_1\A(\lambda)\pa \r(\lambda)\ri)  \d \lambda
\eea
We want to identify this last integral; recall the notation $\Xi^{-1} = \le[{\mathbf C}\atop \mathbf D\ri]$ and that the jumps for $\Xi^{-T}$ imply for the column $\mathbf C^T$
\be
\mathbf C^T(\lambda)  = \le[\1_r \atop 0\ri] +
 \int_{\gamma_+} \frac { \mathbf D^T(\xi) \r^T(\xi)\d\xi}{\xi-\lambda}\label{jump0}
\ee
Moreover, by definition of inverses we have 
\be
\mathbf C (\lambda)\A(\lambda) \equiv \1_r\equiv \mathbf D(\lambda) \B(\lambda).
\ee

Consider 
\bea
 (\mathcal R_{++} \circ\pa \mathcal K)(\lambda, \mu) = \int_{\gamma_+}\mathcal
  R_{++}(\lambda ,\xi)\frac {E_1^T(\xi)\pa E_2(\mu) +\pa  E_1^T(\mu) E_2(\xi)}{\mu + \xi} \d\xi 
  =\nonumber\\
 = 
 \int_{\gamma_+}
 \frac{E_1^T(\lambda) \A^\T(\lambda) \mathbf D^T(\xi)E_2(\xi) }{\lambda - \xi} \frac {E_1^T(\xi)\pa E_2(\mu) + \pa E_1^T(\xi) E_2(\mu)}{\xi + \mu } 
 \d\xi  
 = \nonumber\\
 =
 E_1^T(\lambda) \A^\T(\lambda)  \le(
 \int_{\gamma_+} \frac {\mathbf D^T(\xi)\r^T (\xi)\d \xi}{(\xi+\mu)(\lambda - \xi)}\ri) \pa E_2(\mu) 
+
 \int_{\gamma_+}
 \frac{E_1(\lambda) \A^T(\lambda)\mathbf D^T(\xi)E_2(\xi) }{\lambda - \xi} \frac {\pa E_1^T(\xi) E_2(\mu)}{\xi + \mu } 
 \d\xi  
 =\nonumber\\
 = 
  \frac{E_1^T(\lambda) \A^T(\lambda)}{\mu+\lambda} \le(
 \int_{\gamma_+}\le(   \frac 1{\xi+\mu}  -    \frac 1{ \xi - \lambda}   \ri) \mathbf D^T(\xi)\r^T (\xi)\d \xi \ri) \pa E_2(\mu)+\nonumber \\
+
 \int_{\gamma_+}
 \frac{E_1^T(\lambda) \A^T(\lambda)\mathbf D^T(\xi)E_2(\xi) }{\lambda - \xi} \frac {\pa E_1^T(\xi) E_2(\mu)}{\xi + \mu } 
 \d\xi   
 =\nonumber\\
 =-
\frac{E_1^T(\lambda) \A^\T(\lambda)}{\mu+\lambda} \le(
 \mathbf C^T(\lambda)  
  -
    \mathbf C^T(-\mu) 
  \ri)  \pa  E_2(\mu)+
 \int_{\gamma_+}
 \frac{E_1^T(\lambda) \A^\T(\lambda)\mathbf D^T(\xi)E_2(\xi) }{\lambda - \xi} \frac {\pa E_1^T(\xi) E_2(\mu)}{\xi + \mu } 
 \d\xi=\nonumber \\   
    = -
  \frac{E_1^T(\lambda) \A^\T(\lambda) }{\mu+\lambda} \le(
\mathbf C^T(\lambda) -   \wh \s_1 \mathbf D^T(\mu)    
  \ri)\pa E_2(\mu)  +
 \int_{\gamma_+}
 \frac{E_1^T(\lambda) \A^\T(\lambda)\mathbf D^T(\xi)E_2(\xi) }{\lambda - \xi} \frac {\pa E_1^T(\xi) E_2(\mu)}{\xi + \mu } 
 \d\xi
  =\nonumber\\
  =
  \frac{E_1^T(\lambda) \A^\T(\lambda) \wh \s_1\mathbf D^T(\mu)\pa E_2(\mu)}{\mu+\lambda}  -
  \frac{ E_1^T(\lambda) \pa E_2(\mu) }{\lambda + \mu}+
   \int_{\gamma_+}
 \frac{E_1^T(\lambda) \A^\T(\lambda) \mathbf D^T (\xi)E_2(\xi) }{\lambda - \xi} \frac {\pa E_1^T(\xi) E_2(\mu)}{\xi + \mu } 
 \d\xi \label{notyet}
\eea
(note that we used (\ref{jump0})). Taking the trace we have to set $\mu=\lambda$ and integrate over $\gamma_+$: the last term in (\ref{notyet}) can then be simplified as well 
\be
\int_{\gamma_+}\!\!\!\!\! \d\lambda \int_{\gamma_+}\!\!\!\!\! \d\xi\,\,\,
 \frac{\Tr\le( E_1^T(\lambda) \A^\T(\lambda) \mathbf D^T(\xi)E_2(\xi)\pa E_1^T(\xi)E_2(\lambda)\ri) }{(\lambda - \xi)(\xi + \lambda) } 
  =
 \\
 = 
\int_{\gamma_+}\!\!\!\!\!\! \d\lambda \int_{\gamma_+}\!\!\!\!\! \d\xi \,\,\,
 \frac{\Tr \le(\r^T(\lambda)  \A^\T(\lambda) \mathbf D^T(\xi)E_2(\xi) \pa E_1^T(\xi)\ri)}{(\lambda - \xi)(\xi + \lambda )} 
=
  \nonumber \\
 =
 \Tr\le[ 
 \int_{\gamma_+}\!\!\!\!\!\! \d\lambda \int_{\gamma_+}\!\!\!\!\! \d\xi\,\,\,
 \r^T(\lambda)  \A^\T(\lambda) \frac 1{2\xi} \le(\frac 1{\lambda - \xi} - \frac 1{\lambda + \xi} \ri)\mathbf D^T(\xi)E_2(\xi)  \pa E_1^T(\xi) \
 \ri] = 
 \\
 =- 
 \Tr\le[\int_{\gamma_+}\!\!\!\!  
\frac{\B^T(\xi) - \B^T(-\xi)}{2\xi} \mathbf D^T(\xi)E_2(\xi)  \pa E_1^T(\xi)
 \d\xi\ri] = \\
 =-
\int_{\gamma_+}\!\!\!\!
\frac{\Tr \le((\1_r - \A^T(\xi)\wh \s_1\mathbf D^T(\xi))E_2(\xi)  \pa E_1^T(\xi)\ri)}{2\xi}
 \d\xi
\ee
Taking the trace of (\ref{notyet}) we thus have
\be
\Tr \le(\mathcal R_{++} \circ \pa \K\ri) =-
\overbrace{\int_{\gamma_+} \d\lambda   \frac{\Tr\le (  E_1^T(\lambda) \pa E_2(\lambda) +\pa  E_1^T(\lambda) E_2(\lambda)
\ri)}{2\lambda}}^{=\Tr \pa \K}
\\
+
\Tr\le[  \int_{\gamma_+}\!\!\!\d\xi  \frac{\A^\T(\xi) \wh \s_1\mathbf D^T(\xi)\pa E_2(\xi) E_1^T(\xi) }{2\xi}  
+
\int_{\gamma_+}\!\!\!\!\d\xi
 \frac{\A^\T(\xi)\wh \s_1 \mathbf D^T (\xi)E_2(\xi) \pa E_1^T(\xi) }{2\xi}
  \ri] 
\ee

 Together we thus have 
\be
\Tr \le( \mathcal R_{++} \circ\pa \mathcal K\ri) =-
\Tr \pa \K+ 
    \int_{\gamma_+} \!\!\!\!\!\d\lambda \frac{\Tr \le( 
\A^\T(\lambda) \wh \s_1\mathbf D^T (\lambda )\pa \r^T(\lambda)\ri) }{2\lambda}\ ,
\ee
and therefore
\bea
\int_{\gamma_+  \cup \gamma_-} \Tr \le(\Gamma_{-}^{-1} \Gamma_-' \pa M M^{-1}\ri) \frac {\d \lambda}{2i\pi} 
- \int_{\gamma_+  \cup \gamma_-} \Tr \le(\Xi_{-}^{-1} \Xi_-' \pa M M^{-1}\ri) \frac {\d \lambda}{2i\pi} =\\
= 
2 \Tr (\mathcal R_{++} \circ \pa \mathcal K) 
+
 2 \Tr \pa \mathcal K
\eea

In summary, using Theorem \ref{4.1}, we have
\be
\int_{\gamma_+  \cup \gamma_-} \Tr \le(\Gamma_{-}^{-1} \Gamma_-' \pa M M^{-1}\ri) \frac {\d \lambda}{2i\pi}  = 
\pa \ln \det( \Id_{\gamma_+} - \mathcal K^2)+
  2 \Tr (\mathcal R_{++} \circ \pa \mathcal K) 
+
 2 \Tr \pa \mathcal K
\label{756} 
\ee
On the other hand we now show that the r.h.s. of (\ref{756}) is precisely $2\pa \ln \det (\Id_{\gamma_+}
+
\mathcal K)$ at which point the proof shall be then complete.
To verify this last point  we have (using Theorem \ref{thmXi})
\bea
(\Id_{\gamma_+} 
 + 
 \mathcal K)^{-1} &=& (\Id_{\gamma_+}-\mathcal K^2)^{-1}(\Id_{\gamma_+}
  - 
 \mathcal K)
 \mathop{=}^{\hbox{\tiny Thm. \ref{thmXi}}} (\Id_{\gamma_+}  +\mathcal R_{++})(\Id_{\gamma_+}
  - 
 \mathcal K) \nonumber\\
&=&\Id_{\gamma_+} 
 - 
 \mathcal K + \mathcal R_{++} 
 - 
 \mathcal R_{++} \mathcal K \label{resolventformula}
\eea
from which we can compute the variations of the determinant
\bea
2\pa \ln \det (\Id_{\gamma_+} 
 + 
 \mathcal K) &\& =
 2
  \Tr ( (\Id_{\gamma_+} 
  + 
  \mathcal K)^{-1} \pa \mathcal K)=\nonumber\\
&\& =
2\Tr \le(
\pa \mathcal K  - \mathcal K \pa \mathcal K +  \mathcal R_{++} \pa \mathcal K    - 
  \mathcal R_{++}  \mathcal K \pa \mathcal K
\ri) =\nonumber\\
&\&=- \Tr ((\Id_{\gamma_+} +\mathcal R_{++} )\pa (\mathcal K^2)) 
 + 
 2\Tr (\pa \mathcal K) 
  + 
 2 \Tr ( \mathcal R_{++}\pa \mathcal K)=\nonumber\\
&\& = \pa \ln \det (\Id_{\gamma_+} - \mathcal K^2) 
 + 
2 \Tr ( \mathcal R_{++}\pa \mathcal K)
+
2\Tr (\pa \mathcal K) 
\eea
\QED

When the dependence on the deformation parameter is in the form specified in (\ref{sdep}) then we can write the differentials in terms of formal residues as shown here.
\begin{prop}
\label{formalres}
Suppose that $M(\lambda)$ in (\ref{3.1}) can be written as  $M(\lambda) = {\rm e}^{T(\lambda)} M_0(\lambda) {\rm e}^{-T(\lambda)}$ where $T(\lambda)$ is a polynomial diagonal matrix without constant term in $\lambda$ ($T(0)=0$) whose entries depend on the deformations and such that $T(-\lambda) = \wh \s_1 T(\lambda) \wh \s_1$. Let $\pa$ be the derivative w.r.t. one deformation parameter.   
Then 
\be
\pa \ln \tau_\Gamma &\& = \omega_\Gamma(\pa) =-\frac 12 \res_{\infty} \Tr\bigg(\Gamma^{-1}(\lambda)\Gamma'(\lambda) \pa T(\lambda)\bigg)\d \lambda ,\label{428}\\
\pa\ln \tau_\Xi &\&=  
\omega_\Xi(\pa) =- \res_{\infty} \Tr\bigg(\Xi^{-1}(\lambda)\Xi'(\lambda) \pa T(\lambda)\bigg)\d \lambda
\label{429}
\ee
where the residues are understood as formal residues, or the coefficient of $\lambda^{-1}$ in the expansion at infinity.
\end{prop}

{\bf Proof of Prop. \ref{formalres}.}
The equivalence of the formal residues  (\ref{428}) with the integral representation (\ref{omegaxi})  (or (\ref{omegagamma})) was proven in \cite{BertolaIsoTau} in a more general context, but we recall here the gist of it.
The formal residue in (\ref{428}) (for the case of $\omega_\Gamma$, the other case being completely analogous) can be written as an integral on an  expanding counterclockwise circle (with the piecewise-defined $\Gamma$) and then it  can be transferred by the use of Cauchy theorem to the integral 
\bea
-\frac 12 \res_{\lambda=\infty} \Tr (\Gamma^{-1} \Gamma' \pa T) \d\lambda=
+
\frac 1 2 \lim_{R\to \infty} \oint_{|\lambda|=R}   \hspace{-10pt}
\Tr (\Gamma^{-1} \Gamma' \pa T)\frac{ \d\lambda}{2i\pi}   =\\
= 
\frac 1 2 \int_{\gamma_+\cup \gamma_-} \hspace{-10pt}\Tr \le[\le(
-\Gamma_{+}^{-1} 
\Gamma_{+ }' + \Gamma_{-}^{-1} \Gamma_{-}
' \ri)\pa T\ri]\frac{ \d\lambda}{2i\pi}  
+
\cancel{ \frac 1 2 \oint_{|\lambda|=\epsilon}   \hspace{-10pt}
\Tr (\Gamma^{-1} \Gamma' \pa T)\frac{ \d\lambda}{2i\pi} } 
=\\
=-\frac 1 2 \int_{\gamma_+\cup \gamma_-} \hspace{-10pt}
\Tr \le[\le(  
 M^{-1} \Gamma_-^{-1} \Gamma_-' M +
 M^{-1} M' -  \Gamma_-^{-1} \Gamma_-'  \ri)\pa T\ri]\frac{ \d\lambda}{2i\pi} =\cr
=-\frac 1 2 \int_{\gamma_+\cup \gamma_-} \hspace{-10pt}\Tr \le[
 \Gamma_-^{-1} \Gamma_-' \le(  M \pa T M ^{-1} -\pa T \ri) 
+
  M^{-1} M'  \pa T\ri]\frac{ \d\lambda}{2i\pi} \label{414}
\eea
Firstly, the term crossed out is zero because $T(\lambda)=\mathcal O(\lambda) \ \Rightarrow\ \pa T(\lambda)=\mathcal O(\lambda)$ (as $\lambda\to 0$) and $\Gamma^{-1}\Gamma'$ may have at most a simple pole at $\lambda=0$ (thanks to (\ref{0asy})) so that the term is analytic at $\lambda=0$.
Secondly, note that $M^{-1} M'$ is (piecewise) strictly upper or lower triangular on $\gamma_+\cup \gamma_-$ and $\pa T $ is diagonal, hence the term $\Tr (M^{-1} M' \pa T)\equiv 0$ on $\gamma_+\cup \gamma_-$. On the other hand from  the formula (\ref{sdep}) follows immediately that 
$
M \pa T M^{-1} -\pa T   = -\pa M M^{-1} 
$
and hence (\ref{428}) gives exactly (\ref{omegagamma}).  \QED\
The following corollary follows from direct matrix multiplications using the asymptotic forms (\ref{expGamma}) and (\ref{34}) together with the special structure of the expansion matrices (\ref{314}) and (\ref{419}).
\begin{cor}
\label{corsimple}
If $\pa T(\lambda) =i \lambda\, e_{kk}\ot \s_3$ with $e_{kk}$ the diagonal elementary ($r\times r$) matrix,  then 
\be
\omega_\Gamma (\pa) = -\frac 1 2 \res_{\infty} \bigg(\Gamma^{-1} \Gamma' \lambda e_{kk}\ot \s_3\bigg)\d\lambda =-{i} (a_1)_{k,k}
\ee
Similarly 
\be
\omega_\Xi (\pa) = - \res_{\infty} \bigg(\Xi^{-1} \Xi ' \lambda e_{kk}\ot \s_3\bigg)\d\lambda = -2 i (\a_1)_{k,k}
\ee
\end{cor}

To conclude this section we prove that, when $r=1$ the relation between the two Riemann--Hilbert problems, and in particular equation (\ref{518}), can be interpreted as a Miura transformation between the two tau functions. 
\begin{prop}
\label{taumiura}
Suppose  $r=1$ (but possibly $p\geq 1$) and $M(\lambda):={\rm e}^{is\lambda\s_3}M_0(\lambda){\rm e}^{-is\lambda\s_3}$ (i.e. a special case of Prop. \ref{formalres}). Then the tau functions (Fredholm determinants)  $\tau_\Xi,\tau_\Gamma$ are related through the Miura transformation
	\be\label{Miura}
		\le(\pa_s\ln\tau_{\Xi}-2\pa_s \ln \tau_{\Gamma}\ri)^2 = -\pa_s^2 \ln \tau_{\Xi}
	\ee
Equivalently we may simply write 
\be
\le( \omega _\Xi(\pa_s)  - 2 \omega_\Gamma(\pa_s) \ri) ^2= -\pa_s \omega_\Xi(\pa_s)
\ee	
\end{prop}
{\bf Proof of Prop. \ref{taumiura}.}
The solution $\Xi$ of Problem \ref{prob3} is such that 
\be
\pa_s \le(\Xi(\lambda;s){\rm e}^{is \lambda \s_3}\ri) = U (\lambda;s)\Xi(\lambda;s){\rm e}^{is\lambda s_3}\label{Lax1}\\
U(\lambda;s):= i\lambda \s_3 + 2\b_1(s) \s_1
\ee 
which can be easily proved by noticing that $U$ has no jumps, hence it is entire, and then by looking at the behavior at infinity using the expansion of $\Xi$. On the other hand then comparing the terms in the expansion of the two sides of (\ref{Lax1}) one finds that 
\be
\pa_s \a_1 = -2i \b_1^2\label{3.9}\ .
\ee

Now, Corollary \ref{corsimple} (i.e. Prop. \ref{formalres}) yields
\be
\pa_s\ln \tau_{_\Xi} &=& -\res_{\lambda=\infty} \Tr\le(\Xi^{-1} \Xi' \pa_s T\ri) \d\lambda =-2i\a_1\label{3.6}\\
\pa_s\ln \tau_{_\Gamma} &=& -\frac 12 \res_{\lambda=\infty} \Tr\le(\Gamma^{-1} \Gamma' \pa_s T\ri) {\d\lambda} = -ia_1\label{3.7}
\ee
Rewriting (\ref{518}) as 
$
	2\b_1^2= (2ia_1 -2i\a_1)^2 
$
and using the equations  (\ref{3.9}), (\ref{3.6}), (\ref{3.7}) we obtain the statement of the proposition. \QED
\begin{remark}\label{remMiu}
	Defining $u:=2\pa_s^2\ln\tau_{\Gamma}$ and $v^2:=-\pa_s^2\ln\tau_{\Xi}$ we obtain the usual formulation of the Miura transformation
	$$u=-v^2\pm \pa_s v.$$
\end{remark}

\section{Applications: Fredholm determinants and noncommutative  Painlev\'e II, XXXIV}
\label{ncAirykernel}
We now consider the Fredholm determinant for the convolution operator on $L^2(\R_+, \C^r)$ given by 
\be
(\mathcal Ai_{\vec s}\,f)(x)&\&  := \int_{\R_+} {\mathbf {Ai}}(x+y;\vec s) f(y)\d y\\
{\bf Ai} (x;\vec s)&\& :=\int_{\gamma_+}  {\rm e}^{\theta(\mu)} C{\rm e}^{\theta(\mu)}{\rm e}^{i x\mu} \frac {\d \mu}{2\pi} = \le[c_{jk} \Ai(x+s_j+s_k)\ri]_{j,k}\\
&\& \theta:=  \frac {i\mu^3}6\1_r  + \le[
\begin{array}{ccccc}
 i s_1 \mu &&&\\
& i s_2 \mu \\
&&\ddots\\
&&& i s_r  \mu 
\end{array}
\ri] = \frac {i\mu^3}6 \1 + i \ss \mu\label{5.3} \\
&& \ss :={\rm diag}(s_1, s_2,\dots,s_r)
\ee
The matrix $C$ is a constant $r\times r$ matrix and the contour $\gamma_+$ is a contour contained in the upper half plane and extending to infinity along the directions $\arg(z) = \frac \pi 6, \frac {5\pi}6$.

Here the matrices $E_1,E_2$ can be chosen as 
\be
E_1(\lambda) =- \frac 1{2i\pi} {\rm e}^{\theta(\lambda)} C ,\ \ E_2(\lambda) = {\rm e}^{\theta(\lambda)}
\ ,\ \ \r(\lambda) = -\frac 1{2i\pi} {\rm e}^{\theta(\lambda)} C {\rm e}^{\theta(\lambda)}
\label{ncHMcL}
\ee

The first issue is whether the solutions of Problems \ref{prob1}, \ref{prob3} exist for real values of the parameters $\vec s$.
We shall -in fact- show an existence theorem for Problem \ref{prob3}, which immediately implies existence of the solution of Problem  \ref{prob1} by Proposition \ref{RHPrelation}.

\begin{shaded}
\begin{theorem}
\label{Xiexists}
Suppose $C=C^\dagger$ is a Hermitean matrix; then the solution to Problem \ref{prob3} with $\r$ as in (\ref{ncHMcL}) exists for all values of $\vec s\in \R^r$ if and only if the eigenvalues of $C$ are all in the interval $[-1,1]$. 
If $C$ is an arbitrary complex matrix with {\em singular values}  in $[0,1]$ then the solution still exists for  all $\vec s\in \R^r$.
(The singular values of a matrix are the square roots of the eigenvalues of $C^\dagger C$)
\end{theorem}
\end{shaded}
{\bf Proof of Thm. \ref{Xiexists}}
The proof is based on the estimate of the operatorial norm of the operator $\mathcal Ai_{\vec s}$ (with parameter $C\in \Mat(r\times r,\C)$); the invertibility of the operator $\Id + \mathcal Ai_{\vec s}$ will be guaranteed if the norm of $\mathcal Ai_{\vec s}$ is less than one.

On the other hand the invertibility is equivalent to the non-vanishing of the respective Fredholm determinant; hence from Corollary \ref{cor2.1} we have  (in the present notation, with $p=r$)
\be
\det \le(\Id_{\R_+} \pm\mathcal Ai_{\vec s} \ri)_{L^2(\R_+,\C^r)}   = \det \le(\Id_{\gamma_+} \pm \mathcal K\ri)_{  L^2(\gamma_+,\C^p)} \ \ \Rightarrow\\
\det \le(\Id_{\R_+} - \mathcal Ai_{\vec s}^2  \ri)_{L^2(\R_+,\C^r)}   =\det \le(\Id_{\gamma_+} - \mathcal K^2\ri)_{  L^2(\gamma_+,\C^p)}
\label{57}
\ee
Thus, if $\|| \mathcal Ai_{\vec s}\||<1$ then  $\|| \mathcal Ai_{\vec s}^2\||<1$  and thus the Fredholm determinants on the line (\ref{57}) do not vanish. This is sufficient for the existence of the solution of Problem \ref{prob3} as shown in Thm. \ref{thmXi}. 

Let us then estimate the  norm $\|| \mathcal Ai_{\vec s}\||$; 
first of all note that $L^2(\R_+) \simeq L^2([s,\infty))$ by simple translation; with this in mind we can express the operator $\mathcal Ai_{\vec s}$ as the operator $\mathcal Ai_{\vec 0}$ but acting on the space 
\be
\mathcal H_{\vec s} &\&= L^2([s_1,\infty)) \oplus\dots \oplus L^2([s_r,\infty))\\
\mathcal Ai_{\vec 0}: &\&\mathcal H_{\vec s}  \to \mathcal H_{\vec s} \\
&\&(f_1,\dots f_r) \mapsto \le(\sum_{k=1}^r C_{jk} \int_{\R} \Ai(x+y) \chi_{[s_k,\infty)} f_k(y)\d y\ri)_{j=1,\dots, r}
\ee
Let $P_{\vec s}$ be the orthogonal projector 
\be
P_{\vec s}:L^2(\R,\C^r) \to \H_{\vec s}\ ,\qquad P_{\vec s} = {\rm diag} (\chi_{_{[s_1,\infty)}}, \dots , \chi_{_{[s_r,\infty)}})
\ee
Then we have 
$\mathcal Ai_{\vec s} \simeq P_{\vec s} \mathcal Ai_{\vec 0} P_{\vec s}$. On the other hand it is evident that 
the operator $
\mathcal Ai_{\vec 0} :L^2(\R,\C^r) \to :L^2(\R,\C^r)
$ is the tensor product $ \mathcal Ai_{\vec 0} = C\otimes \mathcal L$
where we have denote $\mathcal L$ the {\em scalar} convolution operator with the Airy function on $\R_+$
\be
\mathcal L:&\&L^2(\R)\to L^2(\R)\\
&\& \mathcal L f(x):= \int_{\R} \Ai(x+y) f(y)\d y
\ee 
This operator squares to the identity (as it is easily seen in Fourier transform, but is also well  known \cite{TracyWidomLevel}) and hence has unit norm (in fact it is a unitary operator, an easily verified fact in Fourier transform).
Therefore 
\be
\|| \mathcal Ai_{\vec 0} \|| = \|| C\otimes \mathcal L\|| = \||C\||  \||\mathcal L\|| = \||\mathcal C\||\ ,
\ee
and hence 
\be
\|| \mathcal Ai_{\vec 0} \||   =\|| P_{\vec s} \mathcal Ai_{\vec 0}P_{\vec s}  \||\leq \|| P_{\vec s} \||\,||| \mathcal Ai_{\vec 0}|||\,|||P_{\vec s}  \|| <   \||\mathcal C\||
\ee
Since $|||C|||$ is the maximal singular value the first part of the theorem is proved because $|||C|||\leq 1$ implies that the norm of our operator is {\em strictly} less than one.

To prove necessity in the case $C=C^\dagger$ Hermitean,  suppose that $C$ has an eigenvalue $\varkappa\in \R\setminus [-1,1]$ with eigenvector $\vec v_0$ (for Hermitean (and more generally normal) matrices the singular values are simply the absolute values of the eigenvalues). We need to show that for some choice of $\vec s$ the Fredholm determinant vanishes; we will accomplish this by finding a special value of $\vec s$ for which the square of $\mathcal Ai_{\vec s}$ has an eigenvector and hence it is not invertible, thus implying the non-solubility of Prob. \ref{prob3}. To this end we take $\vec s = (s,s,s,s\dots,s)$ and $\vec f(y) = \vec v_0 \varphi(y)$ with $\varphi(y)\in L^2(\R_+)$. Then 
\be
(\mathcal Ai_{\vec s}^2 \vec f)(x) =  \kappa^2 \vec v_0 \int_{\R_+} K_{\Ai}(x+s ,y+s ) \varphi(y) \d y
\ee
where $K_{\Ai}$ is the well known Airy kernel
\be
K_{\Ai}(x,y) = \frac {\Ai(x) \Ai'(y)- \Ai'(x) \Ai(y)} {x-y}= \int_{\R_+} \Ai(x+z) \Ai(z+y)\d z
\ee
It is well--known that $K_{\Ai}$ on $L^2([s,\infty))$ is self-adjoint and of trace-class. Let $\Lambda(s)$ be the maximum eigenvalue. This is a continuous function of $s$ and tends to $1$ as $s\to-\infty$ (it clearly tends to zero as $s\to +\infty$) \cite{TracyWidomLevel}. Let $\varphi_s(y)$ be the corresponding eigenfunction and use now $f(y) = \vec v_0 \varphi_s(y)$.
Then 
\be
({\mathcal Ai}_{\vec s}^2 \vec f)(x)  = \Lambda (s)\kappa^2 \vec f(x)
\ee
If $\kappa^2>1$ there is a value of $s_0\in \R$ for which $({\mathcal  Ai}_{\vec s}^2 f)(x)=f(x)$, thus proving that $\Id - \mathcal Ai_{\vec s}^2$ cannot be invertible for $\vec s =(s_0,\dots, s_0)$. This concludes the proof.
 \QED
 The reader may now wonder whether these kernels have a ``physical'' interpretation. The answer is in the affirmative 
 \begin {theorem}
 \label{thmdetpp}
If $C$ is real or Hermitean then the kernel $\mathcal Ai^2_{\vec s}$ is totally positive on the set $\{1,2,\dots r\} \times \R$.
 \end{theorem}
 {\bf Proof}
 It is a general result that if $(X,\d\mu)$ is a measure space  then for any $W\subset X$ we have 
 \be
k! \det\le[\int_W f_a(\zeta) g_b(\zeta) \d\mu(\zeta)\ri]_{a,b=1..k} = \int_{W^k}\det [f_a(\zeta_c)] \det [g_b(\zeta_d)]  \prod_{j=1}^k \d\mu(\zeta_j) \label{debruijn}
 \ee
 In our case we take $X=\{1,\dots,r\}\times \R$ with the counting measure times Lebesgue measure. A function on $X$ is then equivalently interpreted as a vector of usual functions: $\xi=(j,x)$\ $\Rightarrow$ $f(\xi) = f((j,x))=:f_j(x)$. With this understanding the kernel $\Ai^2_{\vec s}$ is understood as a scalar function on $X\times X$ to wit
 \be
 \mathcal Ai_{\vec s}^2(\xi_1,\xi_2) = \\=
 [\mathcal Ai_{\vec s}^2]_{j_1,j_2}(x_1,x_2) = \sum_{k=1}^r c_{j_1k}c_{k j_2} \int_{\R_+} \!\!\!\! \d z\Ai(x_1+z+s_{j_1}+s_k) \Ai(x_2+z+s_{j_2}+s_k)=\\
 =\int_{X_+}\d\mu(\xi) F(\xi_1,\zeta) F(\zeta, \xi_2)
 \ee
where we have set 
\be
F(\xi_1,\zeta):= c_{j_1,k} \Ai(x_1+s_{j_1} +z+ s_k) \\
\ \zeta=(k,z)\in X_+:= \{1,\dots,r\}\times \R_+\subset X
\ee
The statement of total positivity amounts to checking that for any $K\in \mathbb N$ and any $\xi_1,\dots, \xi_K\in X$ we have 
\be
\det\le[\mathcal Ai_{\vec s}^{2}(\xi_a,\xi_b)\ri]_{a,b\leq K}>0
\ee
To this end we use the previous fact (\ref{debruijn}) and we have 
\be
\det \le[ \mathcal Ai_{\vec s}^2(\xi_a,\xi_b)\ri]_{1\leq a,b\leq K} = \det \le[\int_{X_+}\d\mu(\xi) F(\xi_a,\zeta) F(\zeta,\xi_b)\ri]_{a,b} = \\
=\frac 1 {K!} \int_{X_+^K}\!\!\!\! \det \le[F(\xi_a,\zeta_c)\ri]  \det \le[F(\zeta_c,\xi_a)\ri]  \prod_{c=1}^K \d\mu(\zeta_c)= \frac 1 {K!} \int_{X_+^K}\!\!\! \le| \det \le[F(\xi_a,\zeta_c)\ri] \ri|^2 \prod_{c=1}^K \d\mu(\zeta_c)>0
\ee
where the modulus occurs if $C$ is complex Hermitean (in which case $F(\xi,\zeta)= \ov F(\zeta,\xi)$), while if $C$ is any real matrix then we have a simple square (which is anyway positive).
 \QED

 Theorem \ref{thmdetpp} allows us to interpret the kernel $\mathcal Ai^2_{\vec s}$ as defining a {\bf determinantal point process} on the space of configurations $X=\{1,\dots, r\}\times \R$ \cite{Sosh:RandomPointFields}. The Fredholm determinant is then a multi-level gap distribution for said process on the interval $[S,\infty)$ (after a translation $x\mapsto x-S$).
 \subsection{Noncommutative Painlev\'e\ II and its pole-free solutions}
 \label{PIILax}
We consider first Problem \ref{prob3} for $\Xi$; the jump is written ($\r$ defined in \ref{ncHMcL})
\be
\Xi_+ = \Xi_- \le(\1_{2r} - 2i\pi \r(\lambda) \ot  \s_+\ri) \ ,\ \ \lambda\in \gamma_+\\
\Xi_+ = \Xi_- \le(\1_{2r} - 2i\pi \r(-\lambda) \ot  \s_-\ri) \ ,\ \ \lambda\in \gamma_-
\ee

The matrix $\Psi(\lambda):= \Xi(\lambda) {\rm e}^{\theta(\lambda) \ot  \s_3},$ with $\theta(\lambda)$ as in (\ref{5.3}), solves a RHP with constant
jumps
\be
\Psi_+ &\&= \Psi_- \le(\1_{2r} + C \ot  \s_+\ri) \ ,\ \ \lambda\in \gamma_+
\nonumber \\
\Psi_+ &\& = \Psi_- \le(\1_{2r} + C \ot  \s_-\ri) \ ,\ \ \lambda\in \gamma_-
\nonumber \\
\Psi(\lambda) &\&= \le(\1_{2r} + \mathcal O(\lambda^{-1}) \ri) {\rm e}^{\theta(\lambda)\ot \s_3}\ ,\ \ \ \lambda \to\infty
\label{PsiRHP}
\ee
It would be simple to show that $\Psi(\lambda)$ solves a polynomial ODE in $\lambda$ (of degree $2$, see Lemma \ref{nc1}), which eventually would lead to showing that $\beta_1(\vec s)$ solves a noncommutative version of the Painlev\'e\ II equation (whence the title of the section). In this perspective, the above jumps are a {\em particular} choice of  Stokes' multipliers associated to such an ODE, exactly as in the scalar commutative Lax representation of PII \cite{Its:2003p5}.  We thus describe in the next section below, {\em ex ante}, the most general set of generalized monodromy data  for the ODE (\ref{Amatrix}).

 \subsubsection{The general Stokes'  data/Riemann--Hilbert problem for \texorpdfstring{$\Psi$}{wer}}
 Denote by $\Delta\subset \C^r$ the set of diagonals
 \be
 \Delta:=\{\vec s\in \C^r:\ \ s_j=s_\ell,\ \ j\neq \ell\}\ .
 \ee
 Let $\vec s^{(0)} \in \C^r\setminus \Delta$ and choose a ray $\gamma_R:=\R_+{\rm e}^{-\varphi_R}$ in such a  way that $\Im z(s_k^{(0)}-s_\ell^{(0)}) \neq 0$ for $z\in \gamma_R$. Let $\gamma_L:= -\gamma_R$.
 We  introduce the ordering $k\prec \ell$ as follows 
 \be
k \prec \ell \ \  \Leftrightarrow\ \ \Im ({\rm e}^{i\vartheta_R} (s_k^{(0)}-s_\ell^{(0)}))    < 0 
\label{order}
\ee

For a fixed $\gamma_R$ this ordering is constant in a suitable open conical neighborhood of $\vec s^{(0)}$ not intersecting the diagonals $\Delta$ (as should be clear by a simple continuity argument): we shall understand such choice of neighborhood and keep the chosen ordering fixed.
We shall say that a matrix $N$ is {\bf upper(lower)-triangular relative to the ordering (\ref{order})} if $N_{k\ell}=0$ for $k\prec \ell$ ($\ell\prec k$, respectively) and $N_{kk}=0$.
\begin{example}
If $s_1<s_2<\dots s_r$ are real and ordered, then the notion of upper(lower) triangularity  relative to any ray $\arg z =(-\pi,0) $ is the usual one.
\end{example}
We define  the six additional contours
\be
\gamma_j:= \R_+{\rm e}^{\frac {ik\pi}3+ \frac \pi 6}\ ,\ \ k=0,\dots, 5
\ee 

Let $C_0,\dots C_2$  three arbitrary $r\times r$ matrices,  $S_u=\1_r+N_u, S_l=\1_r+N_l$ with $N_u,N_l$ two upper/lower triangular matrices relative to the ordering (\ref{order}) 
 determined by the choice of $\gamma_R$, and let ${\bf M} = {\rm diag}(\mu_1,\dots, \mu_r) \in SL(r,\C)$ (traceless). The entries of ${\bf M}$ will be referred to as {\bf exponents of formal monodromy}.
\begin{problem} 
\label{PsincP2}
Let $\Psi(\lambda)$ be a sectionally analytic function on $\C\setminus \gamma_0\cup\dots\gamma_5\cup \gamma_L\cup \gamma_R$, bounded over compact sets of $\C$ and solving the following Riemann--Hilbert problem 
\be
\Psi_+ &\& = \Psi_-(\1_{2r} + C_{0}\ot \s_+)\ ,\ \ \ \lambda \in \gamma_0\ ,\qquad
\Psi_+ = \Psi_-(\1_{2r} + C_{1}\ot \s_-)\ ,\ \ \ \lambda \in \gamma_1\\
\Psi_+ &\&= \Psi_-(\1_{2r} + C_{2}\ot \s_+)\ ,\ \ \ \lambda \in \gamma_2\  ,\qquad
\Psi_+ = \Psi_-(\1_{2r} + C_{0}\ot \s_-)\ ,\ \ \ \lambda \in \gamma_3\\
\Psi_+ &\&= \Psi_-(\1_{2r} + C_{1}\ot \s_+)\ ,\ \ \ \lambda \in \gamma_4\ ,\qquad
\Psi_+ = \Psi_-(\1_{2r} + C_{2}\ot \s_-)\ ,\ \ \ \lambda \in \gamma_5 \\
\Psi_+&\& = \Psi_-\le[(\1_r + S_u)\oplus (\1_r+S_l)\ri]\ ,\ \ \lambda \in \gamma_R\\
\Psi_+&\& = \Psi_-\le[(\1_r + S_l)\oplus (\1_r+S_u)\ri]\ ,\ \ \lambda \in \gamma_L\\
\Psi_+&\& = \Psi_-{\rm e}^{ - i\pi {\bf M}\ot\1_2}\ ,\ \ \ \lambda \in \R_\pm\\
\Psi(\lambda) &\& =  \le(\1_{2r} + \mathcal O(\lambda^{-1})\ri)
{\rm e}^{-i\epsilon \pi {\bf M}\ot \1_2}
\lambda^{ {\bf M}\ot \1_2} {\rm e}^{\theta(\lambda)\ot \s_3}\ ,\ \  \lambda\to\infty \label{asymptpsi}
\ee
where in (\ref{asymptpsi})  $\epsilon=1$ in the upper half-plane, $\epsilon=0$ in the lower half-plane and $\arg(\lambda)\in [-\pi,\pi)$.
The matrices  $C_0,\dots, C_r, S_u, S_l, {\bf M}$ are chosen satisfy the {\bf no-monodromy} condition stating that the product of the jumps  is the identity. (We choose $\gamma_R\neq \R_\pm$ and all the rays are oriented towards infinity).
\end{problem}
Since the rays $\gamma_L,\gamma_R$ may lie in between different $\gamma_j$'s depending on the value of $\vec s$ the no-monodromy condition may take different forms. For example, if $\vec s\in \R^r\setminus \Delta$ we can choose $\arg (\gamma_R)=\frac \pi 2+\epsilon$ and the no-monodromy condition takes the form  
\be
\le[
\begin{array}{cc}
\1 + S_u & 0 \\
0 & \1 + S_l
\end{array}
\ri](\1_{2r} + C_0\ot \s_+)(\1_{2r} + C_2\ot \s_-)(\1_{2r} + C_1\ot \s_+){\rm e}^{-i\pi {\bf M}}\times\nonumber \\
\times
\le[
\begin{array}{cc}
\1 + S_l & 0 \\
0 & \1 + S_u
\end{array}
\ri](\1_{2r} + C_0\ot \s_-)(\1_{2r} + C_2\ot \s_+)(\1_{2r} + C_1\ot \s_-){\rm e}^{-i\pi {\bf M}}=\1_{2r}
\label{nomonodromy}
\ee

\br
\label{5.1}
The problem associated to the Fredholm determinant  of the operator as in Thm. \ref{Xiexists} corresponds to the particular choice $S_u=S_l = C_1={\bf M}=0$ and $C_0=C=-C_2$.
\er
Note that the jumps satisfy the symmetry $M(-\lambda) = \wh \s_1 M(\lambda) \wh \s_1$ and hence we also have (noticing that $\theta(-\lambda)\ot \s_3 = \wh \s_1 \theta(\lambda)\ot \s_3 \wh \s_1$ since $\theta(-\lambda)=-\theta(\lambda)$ as per (\ref{5.3})) 
\be
\Psi(-\lambda) = \wh \s_1\Psi(\lambda)\wh \s_1
\ee
The dimension of the manifold $(C_0,C_1,C_2,S_u,S_l, {\bf M})$ of solutions of (\ref{nomonodromy}) can be computed by noticing that there are a total of $3r^2 + 2 \frac {r (r-1)}2 + r-1=4r^2-1$ variables. The equation (\ref{nomonodromy}) is of the form $A\wh \s_1 A\wh \s_1=\1_{2r}$ and hence --due to the symmetry of conjugation by $\wh \s_1$-- there are only $2r^2$ independent equations. Of these, one is redundant since the determinant of $A$ is already unit. Hence there are $2r^2-1$ independent equations and thus the manifold of solutions has dimension $2r^2$. 
\begin{lemma}
\label{nc1}
Let the matrix $\Psi(\lambda)$ be the solution of the Problem \ref{PsincP2} and  denote the asymptotic expansion at $\infty$ as 
\be
\Psi(\lambda)&\& 
{\rm e}^{i\epsilon \pi {\bf M}\ot \1_2}
\lambda^{-{\bf M}\ot \1_2}{\rm e}^{-\theta(\lambda) \ot \s_3}  = \1_{2r} +\sum_{j=1}^\infty\frac{ \Xi_j}{ \lambda^{j}}\ ,\nonumber \\
&\&\ \ \Xi_{2j+1} = \a_{2j+1} \ot \s_3 + \b_{2j+1} \s_2\ ,\ \ \Xi_{2j} = \a_{2j} \ot \1 + \b_{2j} \s_1\ ,
\label{asymptPsi}
\ee
(recall that $\epsilon=1$ for $\Im \lambda>0$ and $\epsilon=0$ for $\Im \lambda<0$)
 where the expansion is valid sectorially and independent of the sector. Then 
\be
\pa_{s_j}  \Psi &\&= U_j \Psi\ ,\ \ \ U_j = i\lambda\, {\mathrm e}_{j} \ot  s_3 +i [\a_1,{\mathrm e}_{j}] \ot\1 + \{\b_1,{\mathrm e}_{j}\} \ot  \s_1\\
\pa_\lambda \Psi &\&= A(\lambda)\Psi\ ,\ \ 
A(\lambda) = \frac \lambda 2  \sum_{j=1}^r U_j -\frac 1 2 \D \le(\a_1 \ot  \s_3 + \b_1 \ot  \s_2\ri) + i \ss\ot \s_3=\\
&\&A(\lambda)= i\frac {\lambda^2}2 \wh \s_3 + \lambda \b_1\ot \s_1 - \frac 1 2 \D \b_1 \ot  \s_2 +i(\b_1^2 + \ss)\ot  \s_3
\label{Amatrix}
\ee
where
\be
\D:= \sum_{j=1}^r \pa_{s_j}\ ,\ \ {\mathrm e}_{j}:={\rm diag}(0,0,\dots, 1,0,\dots) 
\ee
with the one in the $j$-th position.
\end{lemma}
{\bf Proof}
The fact that the expansion for  (recall that $\arg(\lambda)\in [-\pi,\pi)$)
\be
\Xi(\lambda):= \le\{
\begin{array}{cc}
\Psi(\lambda) {\rm e}^{i \pi {\bf M}}  \lambda^{-{\bf M}\ot \1_2}{\rm e}^{-\theta(\lambda)\ot \s_3} &\Im \lambda>0\ ,\ \  |\lambda|>1\\
\Psi(\lambda) \lambda^{-{\bf M}\ot \1_2} {\rm e}^{-\theta(\lambda)\ot \s_3} &\Im \lambda<0\ ,\ \  |\lambda|>1\\
 \Psi(\lambda) & |\lambda|<1
\end{array}
\ri.
\ee
near $\lambda = \infty$ is of the form in (\ref{asymptPsi}) follows from the symmetry $\Psi(-\lambda)= \wh \s_1 \Psi(\lambda)\wh \s_1$ which then implies the same symmetry for $\Xi$.
The function $\Xi$ has then no jumps on $\R_\pm\setminus \{|\lambda|<1\}$ and the remaining jumps are those of $\Psi$ conjugated by ${\rm e}^{-i\epsilon{\bf M}}\lambda^{{\bf M}\ot \1_2} {\rm e}^{\theta(\lambda)\ot \s_3}$. 
The fact that the expansion is independent of the sector is a consequence of the fact that the jumps for $\Xi$ along the eight rays are analytic in a small open sector around said rays and of the form 
\be
\Xi_+(\lambda) = \Xi_-(\lambda) (\1 + \mathcal O(\lambda^{-\infty})) \ ,\ \ \lambda\to \infty, \ \ \lambda \in \gamma_0\cup \dots \gamma_5\cup \gamma_L\cup \gamma_R
\ee 
uniformly within said sectors.
The fact that $U_j$ and $A$ are polynomials is an immediate consequence of the fact that the jumps of $\Psi$ are independent of $\lambda,\vec s$.
Using Liouville's theorem and the fact that $\pa_{s_j} \Psi \Psi^{-1}$ is entire (a simple consequence of the independence on $s_j$ of the jumps) 
we deduce immediately that $U_j(z)$ can only be a polynomial of degree $1$.
Then 
\be
\pa_j \Xi(\lambda) + i \lambda  \Xi(\lambda) {\mathrm e}_{j} \ot  \s_3 = \le(U^{(1)}_j \lambda + U^{(0)}_j \ri) \Xi(\lambda )\nonumber \\
\frac {\pa_j(\a_1\ot \s_3 + \b_1\ot \s_2) }\lambda  +\dots  + i \lambda  \le(\1 + \frac {\a_1\ot \s_3 + \b_1\ot \s_1 }\lambda + \frac {\a_2\ot \1 + \b_2\ot \s_1}{\lambda^2}+\dots \ri){\mathrm e}_{j}\ot \s_3 =\nonumber \\=
 \le(U_j^{(1)} \lambda + U_j^{(0)} \ri)\le(\1 + \frac {\a_1\ot \s_3 + \b_1\ot \s_1 }\lambda + \frac {\a_2\ot \1 + \b_2\ot \s_1}{\lambda^2}+\dots \ri)\label{917_1}
\ee
Comparing the coefficients of the powers of $\lambda$ we have 
\be
&& \lambda:\ \ \Rightarrow\ \ U_j^{(1)} = i\lambda {\mathrm e}_{j}\ot \s_3 \\
&& \lambda^0:\ \ \Rightarrow\  i\le[ \a_1\ot \s_3 + \b_1\ot \s_2, {\mathrm e}_{j}\ot \s_3\ri] =  U_j^{(0)}\\
&&\lambda^{-1}:\ \ \Rightarrow  \pa_{s_j} \le(\a_1\ot \s_3 + \b_1 \ot \s_3\ri) =- i \le[
\a_2\ot \1 + \b_2 \ot \s_1, {\mathrm e}_{j}\ot \s_3
\ri] + U_j^{(0)} \le(\a_1 \ot \s_3 + \b_1\ot \s_1\ri)\cr
&&  \pa_{s_j} \le(\a_1\ot \s_3 + \b_1 \ot \s_3\ri) =- i \le[
\a_2\ot \1 + \b_2 \ot \s_2, {\mathrm e}_{j}\ot \s_3
\ri] + U_j^{(0)} \le(\a_1 \ot \s_3 + \b_1\ot \s_2\ri)
\ee
If we sum up for $j=1,\dots, r$ we obtain the differential equation
\be
\D \le(\a_1\ot \s_3 + \b_1\ot \s_2\ri) &\&= 
-i \le[\a_2\ot \1 + \b_2 \ot \s_2, \1 \ot \s_3 \ri] + i \le[\a_1\ot \s_3,+ \b_1\ot \s_2, \1 \ot \s_3 \ri] (\a_1 \ot \s_3 + \b_1 \ot \s_2)
=\nonumber
\\= 
-2 \b_2 \ot \s_1 &\& +  2 \b_1\ot \s_1 \le(\a_1\ot \s_3 + \b_1\ot \s_2\ri) = 
-2\b_2\ot \s_1 + 2i \b_1\a_1 \ot \s_2 - 2i \b_1^2 \ot \s_3\label{9113}
\ee
In particular 
\be
\D\a_1  = -2i \b_1^2.\label{alphabeta}
\ee
If we look also at the $\lambda^{-2}$ coefficient we find 
\be
\D\b_1 = 2i\b_2 - 2\b_1\a_1 \ , \ \ \ \b_2 = -\frac i 2 \D \b_1 -i \b_1\a_1.\label{beta2}
\ee
Exactly as before we argue that $A(z)$ is a polynomial of degree $2$.
Then we compute 
\be
\pa_\lambda \Xi(\lambda) + \Xi(\lambda) \le(i\frac {\lambda^2}2\1_{r}  +i \ss  \ri)\ot \s_3 = \le(A_2 \lambda^2 + A_1 \lambda +A_0\ri) \Xi(\lambda)\\
\dots +\le(\1 + \frac {\a_1\ot\s_3 + \b_1\ot \s_2}\lambda + \frac {\a_2\ot \1+\b_2\ot \s_1}{\lambda^2}+\dots \ri)\le(i\frac {\lambda^2}2\1_{r}  +i \ss  \ri)\ot \s_3 =
\\
=\le(A_2 z^2 + A_1 z +A_0\ri) \le(\1 + \frac {\a_1\ot \s_3 + \b_1\ot \s_2}z + \frac {\a_2\ot \1 + \b_2\ot \s_1}{z^2}+\dots \ri)\label{917_2}
\ee
Collecting the coefficients
\be
&\& \lambda^2:\ \ \ \Rightarrow \ \  A_2 =  \frac i  2  \lambda^2\1\ot \s_3\\
&\& \lambda^1:\ \ \ \Rightarrow \ \ A_1 =\frac i 2  [\a_1\ot\s_3 + \b_1\ot \s_2,\1\ot \s_3] \\
&\& \lambda ^0:\ \ \ \Rightarrow \ \ 
A_0 = i\ss\ot \s_3 + \frac i 2  [ \a_2\ot \1 + \b_2\ot \s_1,\1\ot \s_3]  - A_1\le(\a_1\ot \s_3 + \b_1\ot \s_2\ri)=
\\
&\&A_0 =i\ss\ot \s_3 -\frac 1 2  \D \le(\a_1\ot \s_3 + \b_1\ot \s_2\ri)
\ee
where we have used  formula (\ref{9113}). The second expression for $A(\lambda)$ (\ref{Amatrix}) follows from (\ref{alphabeta}). The rest of the proof is a simple computation.\QED

\begin{lemma}
\label{lemmancP2}
Let $\Psi$ be as in Lemma \ref{nc1} and denote by $\beta_1=\beta_1(\vec s)$ the $r\times r$ coefficient matrix in $\Xi_1 = \a_1\ot \s_3 + \b_1 \ot \s_2$ of the expansion as in the mentioned Lemma.
Then the matrix function $\b_1(\vec s)\in Mat(r\times r,\C)$ satisfies the noncommutative Painlev\'e\ II equation.
\be
\D^2 \b_1 = 4\{\ss, \b_1\}  + 8 \b_1^3\ ,\label{1_ncP2}\  \ \ 
\ss:= {\rm diag}(s_1,\dots, s_r)\ , \ \ \D:=\sum_{j=1}^r \frac {\pa}{\pa s_j}\ ,\ \ \{X, Y\}= XY+YX
\ee
\end{lemma}
{\bf Proof.}
We use the zero curvature equations
\be
\D \Psi = \sum_{j=1}^rU_j  \Psi =: U_\D \Psi\ ,\qquad
\pa_\lambda \Psi = A \Psi
\label{ncLaxpair}
\\
(\pa_\lambda  U_\D  + U_\D A - \D A + AU_\D) \equiv 0
\ee
with the $U_j$'s introduced in Lemma \ref{nc1}.
We have 
\be
U_\D =i  \lambda \1 \ot \s_3 + 2 \b_1 \ot\s_1\ ,\ \ A = \frac \lambda 2 U_\D - \frac 1 2 \D\le(\a_1\ot \s_3 + \b_1\ot \s_2\ri) + i\ss\ot \s_3\\
A=
\frac i2  \lambda^2 \1 \ot \s_3 + \lambda \b_1\ot \s_1 - \frac 1 2 \D\le(\a_1\ot \s_3 + \b_1\ot \s_2\ri) + i\ss\ot\s_3
\ee
We now compute this expression (for simplicity we denote with a prime the action of $\D$, noting that $\D\ss = \1_r$)
\be
\pa_\lambda  U_\D &\&= i\1\ot \s_3\\
\D A &\&= \lambda    \b_1'\ot  \s_1 -\frac 1 2  \le(\a_1'' \ot \s_3 + \b_1'' \ot \s_2\ri) +i  \1 \ot \s_3\\[1pt]
[U_D,A] &\&= 
\le[
i  \lambda \1 \ot \s_3 + 2 \b_1 \ot\s_1,
- \frac 1 2 \le(\a_1'\ot \s_3 + \b_1'\ot \s_2\ri) + i\ss\ot\s_3
\ri]=
\\
&\&=
  \lambda  \b_1' \ot \s_1 - 2i \b_1\a_1' \ot \s_2  +i \{\b_1 ,\b_1'\} \ot\s_3   - 2  \{\b_1,\ss\}\ot \s_2   
\ee
Hence 
\be
0\equiv&\& \pa_\lambda  U_\D - \D A + [U_\D,A]=\\
=&\&
\frac 1 2 a_1'' \ot \s_3 + \frac 1 2 \b_1'' \ot \s_2- 2i \b_1\a_1' \ot \s_2  +i \{\b_1 ,\b_1'\} \ot\s_3   - 2  \{\b_1,\ss\}\ot \s_2
   \ee
   Using now (\ref{alphabeta}) we have $\a_1'' =-2i \{\b_1,\b_1'\}$ and hence we are left only with 
   \be
  \le( \frac 1 2 \b_1'' -4 \b_1^3      - 2  \{\b_1,\ss\}\ri)\ot \s_2\equiv 0
   \ee
\QED

Thus Lemma \ref{nc1} and the matrices (\ref{ncLaxpair}) provide a Lax matrix representation for the {\em general} solution of the noncommutative Painlev\'e\ equation (\ref{ncP2}) which is parametrized by the $2r^2$ initial values $\b_1(\vec s)$ and $\D \b_1(\vec s)$ at any point $\vec s$. 

It should also be clear that any solution $\b_1$ of the noncommutative Painlev\'e\ II equation (\ref{ncP2}) yields a compatible Lax pair $A(\lambda;\vec s)$ (\ref{Amatrix}) and $U_\D(\lambda;\vec s)$ (\ref{ncLaxpair}); the Stokes' phenomenon (generalized monodromy data) for the ODE $\pa_\lambda\Psi = A\Psi$ can be seen to be given exactly by the data  specified in Problem \ref{PsincP2} (we refer to \cite{Wasow} for the general theory of Stokes' multipliers). Thus any solution of (\ref{ncP2}) is obtained via the above Lax-pair.

\br
The generic solution of $\b_1(\vec s)$ of the noncommutative PII equation will have non-movable singularities on the diagonals $\vec s\in \Delta$; this is due to the presence --in general-- of nontrivial Stokes multipliers along the rays $\gamma_R, \gamma_L$. If those multiplier are trivial  as well as the exponents of formal monodromy i.e. ${\bf M}=0$,  $S_u = 0= S_l$ then those singularities will be absent. In this case the no-monodromy condition (\ref{nomonodromy}) can be spelt out more clearly as 
\be
[C_0,C_1]= 0\ ,\ \ [C_1,C_2]=0,\ \ [C_0,C_2] = 0 \\
C_0+C_2+C_1 + C_0C_1 C_2 = 0
\label{regularP2}
\ee
which yields  a manifold of dimension $r^2 +r$ (the matrices can be generically diagonalized simultaneously and a simple counting yields this number).
The condition (\ref{regularP2}) resembles very closely the ordinary situation $r=1$ of the commutative Lax representation for PII \cite{Its:2002p9}\footnote{Their matrix $\Psi(\lambda)$ has  the symmetry $\Psi(-\lambda) = \s_2\Psi(\lambda)\s_2$, which means that it should be compared with ours after conjugation by ${\rm e}^{i\frac \pi 4 \s_3}$.}.
\er

The importance of the isomonodromic representation for the noncommutative Painlev\'e\ II equation  is that it implies automatically the {\bf Painlev\'e\ property} \cite{Palmer:Zeros} that the only singularities of the solution are poles except -possibly- the singularities on the diagonal manifold $\vec s\in \Delta$ if the Stokes' matrices $S_u,S_l$ are nonzero.

\br
\label{remPs}
Another important remark is that the solution $\b_1(\vec s)$ as a function of the barycentric variables 
\be
S:= \frac 1 r \sum_{j=1}^r s_j \ ,\ \ \ \delta_j = s_j-S
\ee
has only poles as a function of $S$ (note that $\D  = \pa_S$) if $\vec s\not \in \Delta$; this is so because changing $S$ does not change the differences between the $s_j$'s and hence never crosses the diagonal manifold.
\er

\br
To our knowledge, the noncommutative Painlev\'e\ equation (\ref{1_ncP2})  has appeared first in the recent \cite{RetakhRubtsov} where the authors construct special rational solution using the theory of quasi-determinants \cite{Gelfand2RetakhWilson}. Previously, a version with scalar independent variable (hence replacing the anti-commutator by simply $s \b_1$) was studied in \cite{BalandinSokolov}, where the Painlev\'e\ test was applied. It seems that the Lax representation for the noncommutative version of \cite{RetakhRubtsov} appears in the present manuscript for the first time. 
It seems possible to generalize the Lax-pair representation by allowing a pole at $\lambda=0$ in the Lax matrix $A(\lambda)$ (exactly as in the scalar case). For example the compatibility of the following two Lax matrices
\be
A(\lambda) &\&= \frac {i\lambda^2}2 \wh \s_3 + \lambda \b_1\ot\s_1 + i\le(\ss+\b_1^2\ri)\ot \s_3- \frac 12 \b_1'\ot \s_2 + \frac 1 \lambda \Theta \wh \s_1\ ,\qquad U_\D= i \lambda \wh \s_3 + 2\b_1 \ot \s_1\\
&\& \le[\pa_\lambda - A(\lambda), \D - U_\D(\lambda)\ri]=0
\ee
with $\Theta$ an arbitrary scalar (i.e. commutative symbol). The zero-curvature equations are easily verified to yield
\be
\D^2 \b_1 = 4 \{\ss, \b_1\} + 8 \b_1^3 - 4 \Theta
\ee
which is precisely (with different symbols) the Painlev\'e\ II equation studied in \cite{RetakhRubtsov}. From the isomonodromic method, however, the above equation appears to be not the most general that one may obtain by allowing a pole in $A(\lambda)$. We do now dwell further into the matter since it is peripheral to the focus of the present paper.
\er
 
The compatibility equations for the operators $\pa_{s_j} -U_j$, $\pa_{s_k} -U_k$ and $\pa_\lambda  - A$ yield additional equation listed in the Corollary below, which is proved along the same lines (but we will not report the proof here since it is unnecessarily long, straightforward and anyway this has no bearing for our goals)
\begin{cor}
The matrices $\b_1,  \a_1$ satisfy the systems ($\pa_j = \pa_{s_j}$, $\D = \sum \pa_j$, $j=1,\dots,r$)
\be
\pa_j \b_1 =&\&  \frac 1 2 \{{\mathrm e}_{j},\D \b_1\}- i \{{\mathrm e}_{j}, \b_1 a_1\} + i {\mathrm e}_{j}[ \a_1,\b_1] + i \a_1{\mathrm e}_{j}\b_1 + i \b_1{\mathrm e}_{j}\a_1\\
\frac 1 2 \pa_j\D \b_1= &\&  i (\pa_j \b_1) \a_1+
 \{\{\b_1, {\mathrm e}_{j}\} ,\ss\}   
 - \frac i 2 [{\mathrm e}_{j}, \a_1 ]\D \b_1 +\big[ [\a_1, {\mathrm e}_{j} ],\b_1\big] \a_1 +\nonumber \\
&\& + \b_1\{ {\mathrm e}_{j} \b_1\} \b_1   - \frac i 2 \{\D \b_1 \a_1, {\mathrm e}_{j}\}   + \{{\mathrm e}_{j} ,\b_1^3\}  \\
&\&
\1 - {\mathrm e}_{j} + i \bigg[\ss ,[\a_1, {\mathrm e}_{j}  ]\bigg]     +  \frac 1 2 \bigg[ [\b_1 ,\D \b_1], {\mathrm e}_{j} \bigg]  =0
 \ee
 \be
 \{{\mathrm e}_{j} , \pa_k\b_1\} - \{{\mathrm e}_{k} , \pa_j  \b_1\}   =&\&  i[ \b_1 {\mathrm e}_{k}, \a_1 {\mathrm e}_{j}]
 + i [{\mathrm e}_{j} \b_1, {\mathrm e}_{k} \a_1 ] 
 + i {\mathrm e}_{k} [\b_1 ,\a_1] {\mathrm e}_{j}  +\nonumber \\
+ i {\mathrm e}_{j} [\a_1&\& , \b_1] {\mathrm e}_{k}   
 + i [\a_1 {\mathrm e}_{k}, \b_1 {\mathrm e}_{j} ]  
 + i [{\mathrm e}_{j} \a_1 ,{\mathrm e}_{k} \b_1]\ , 
 \\
i[ {\mathrm e}_{k},\pa_j\a_1]  - i [{\mathrm e}_{j} ,\pa_k \a_1] 
=&\&  {\mathrm e}_{k} \a_1^2 {\mathrm e}_{j}-{\mathrm e}_{j} \a_1^2 {\mathrm e}_{k} +{\mathrm e}_{k} \b_1^2 {\mathrm e}_{j}-{\mathrm e}_{j}\b_1^2 {\mathrm e}_{k}
+\nonumber 
\\
+ [{\mathrm e}_{k} \b_1&\&, {\mathrm e}_{j} \b_1] 
+ [\b_1 {\mathrm e}_{k} ,\b_1 {\mathrm e}_{j}] 
+ [{\mathrm e}_{j} \a_1,  {\mathrm e}_{k} \a_1]
+ [\a_1 {\mathrm e}_{j} , \a_1 {\mathrm e}_{k}]\ .
\ee
\end{cor}

\subsection{Pole-free solutions of noncommutative Painlev\'e\ II and Fredholm determinants}
We now return to the specific situation of the RHP associated to the integrable kernel ${\mathcal Ai}_{\vec s}^2$; this is the special case of the setting as explained in Remark \ref{5.1}.

\begin{shaded}
\begin{theorem}
\label{thmncHML}
Let $\Xi=\Xi(\lambda;\vec s)$ be the solution of Problem \ref{prob3} with $\r$ as in (\ref{ncHMcL}); let 
\be
\b_1(\vec s ):= -i\lim_{\lambda\to\infty} \lambda \Xi_{12}(\lambda;\vec s)
\ee
where $\Xi_{ij}$ denote the $r\times r$ blocks of $\Xi$, $i,j=1,2$.
The matrix function $\b_1(\vec s)\in \Mat(r\times r,\C)$ satisfies the noncommutative Painlev\'e\ II equation.
\be
\D^2 \b_1 = 4\ss \b_1 +4 \b_1\ss  + 8 \b_1^3\ ,\label{ncP2}\  \ \ 
\ss:= {\rm diag}(s_1,\dots, s_r)\ , \ \ \D:=\sum_{j=1}^r \frac {\pa}{\pa s_j}
\ee
The asymptotic behavior of the particular solution associated to the Problem \ref{prob3} is as follows: 
if $S:=\frac 1 r \sum_{j=1}^r s_j\ \to+\infty$ and $\delta_j:= s_j-S$, $j=1,\dots, r$ are kept fixed, $|\delta_j|\leq m$,   then
\be
 [\b_1]_{k \ell} =  -c_{k\ell} \Ai (s_k+s_\ell) +  \mathcal O\le(\sqrt{S} {\rm e}^{-\frac 4 3(2S-2m)^\frac 32}\ri) \label{441}
 \ee
If $C$ is Hermitean then so is  the solution $\b_1(\vec s)$ of the noncommutative Painlev\'e\ equation (\ref{ncP2})  and it
is pole-free for all $\vec s\in \R^r$  if and only if the eigenvalues of $C$ are within $[-1,1]$. If $C$ is arbitrary and its singular values lie in $[0,1]$ then the solution is also pole free for $\vec s\in \R^r$.

Finally, the Fredholm determinant $\tau(\vec s):= \det \le(\Id - {\mathcal Ai_{\vec s}}^2\ri)$  satisfies 
\be
 \det \le(\Id - {\mathcal Ai_{\vec s}}^2\ri) = \exp\le( - 4 \int_S^\infty (t-S) \Tr (\b_1^2(t+ \vec \delta)) \d t\ri)
 \ee
 where $t+\vec \delta := (t+\delta_1, \dots, t+\delta_r)$.
\end{theorem}
\end{shaded}
The last statement is the noncommutative (matrix) equivalent of the celebrated Tracy--Widom distribution \cite{TracyWidomLevel}.
Before giving the proof of Thm. \ref{thmncHML} we prove the uniqueness of the solution.
\begin{prop}
\label{HMunique}
For any $r\times r$ matrix $C$ there is a unique solution of noncommutative PII (\ref{ncP2}) with the asymptotics (\ref{441}).
\end{prop}
{\bf Proof of Prop. \ref{HMunique}.}
The proof does not differ significantly from the scalar case as in \cite{Hastings-McLeod}. In barycentric and relative coordinates $S, \delta_j$ as in Thm. \ref{thmncHML} we have  $\D = \pa_S$.
The regime we consider is  $S\to+\infty$ and all $\delta_j$ bounded below.
We note that the function 
\be
[U(\vec s)]_{k\ell}:=- c_{k\ell} \Ai(s_k+s_\ell) =- c_{k\ell} \Ai(2S + \delta_k+\delta_\ell) 
\ee
is a solution of the linear part of (\ref{ncP2}):
\be
\D^2 U_{k\ell} = -4 c_{k\ell} \Ai''(s_k+s_\ell) = -4c_{k,\ell} (s_k+s_\ell) \Ai(s_k+s_\ell) = 4( s_kU_{k\ell} + U_{k\ell}s_\ell).
\ee 
Then any  solution $(\b_1)_{k\ell}$ with the specified asymptotic also solves the integral equation\footnote{We recall that $Bi$ is the solution of $f''=xf$ such that $\mathrm{Wr}(\Ai,\mathrm{Bi})=\pi^{-1}$.}
\be
[\b_1]_{k\ell} = U_{k\ell}+4 \pi\!\!\!\int_{S}^\infty \!\!\! \!\!\le(
\Ai(2S\!+\!\delta_k\!+\!\delta_\ell) \mathrm{Bi}(2t\!+\!\delta_k\!+\!\delta_\ell)\! -\! 
\Ai(2t\!+\!\delta_k\!+\! \delta_\ell) \mathrm{Bi}(2S\!+\!\delta_k\!+\!\delta_\ell) \ri) [(\b_1)^3]_{k\ell}\d t\label{integral}
\ee
Equation (\ref{integral}) can be solved by iterations for $S$ sufficiently large, as noted in \cite{Hastings-McLeod} for the scalar case.

The local uniqueness follows from the local uniqueness of the solution of the ODE (\ref{ncP2}) (in $S$).  
The solution is easily seen to be locally analytic in $\vec s$ because of the analyticity of the ODE and also from the integral equation.
We also point out that since the generalized monodromy data associated to this solution have $S_u=S_l=0={\bf M}$ (see Problem \ref{PsincP2}) then there are no critical singularities at all in $\vec s$ and we may only have poles at most (the Fredholm determinant is analytic in $\vec s$, hence can only have zeroes). Thus the solution is globally defined for $\vec s\in \R$ by analytic continuation.
\QED
\br
Because of (\ref{441}) and Prop. \ref{HMunique}  we may call the special solution of noncommutative PII arising  above {\bf the noncommutative Hastings-McLeod} solution(s). 
\er
\noindent {\bf Proof of Thm. \ref{thmncHML}.}
The fact that $\b_1$ solves the noncommutative PII equation (\ref{ncP2}) follows from Lemma \ref{lemmancP2} since this is a special case of that with $S_u=S_l=C_1={\bf M}=0$ and $C=C_0=-C_2$.

\noindent  {\bf Asymptotics.}
Suppose that $S=\frac 1 r \sum s_j$ is large and positive and $\delta_j:= s_j-S$ are bounded by -let's say- $m$.
We rewrite the RHP in the scaled variable $z:= \frac { \lambda}{\sqrt S}$.  
 The jump on the  contours $\gamma_\pm$ of the form $\1 - 2i\pi \r\ot \s_+,\1-2\pi\tilde\r\ot\s_-$ can be factored into (commuting) matrices (here below $e_{k\ell}$ is the elementary matrix)
\be
\1 - 2i\pi \r \ot \s_+  = \prod _{k,\ell=1}^r \le( \1 +  c_{k\ell} {\rm e}^{ i S^{\frac 32} \le( \frac 1 3 z^3  +(2 + \frac {\delta_k+\delta_\ell}{S} )z\ri)} e_{k,\ell} \ot \s_+ \ri)\label{factorization}\\
\1 - 2i\pi \r \ot \s_-  = \prod _{k,\ell=1}^r \le( \1 +  c_{k\ell} {\rm e}^{ - i S^{\frac 32} \le( \frac 1 3 z^3  +(2 + \frac {\delta_k+\delta_\ell}{S} )z\ri)} e_{k,\ell} \ot \s_- \ri)
\ee
Each factor has a saddle point at $z = \pm i\sqrt{2 + \frac {\delta_k+\delta_\ell}{S} }$ and the contours $\gamma_\pm$ supporting the single jump can be split according to the factorization (\ref{factorization}) so that each of the factor is supported on a different contour $\gamma_\pm^{(k,\ell)}$ of steepest descent for the corresponding phases. Proceeding this way the reader realizes that each factor 
\be
M^{(k,\ell)}_\pm (\sqrt S z)  = \1 +  c_{k\ell} {\rm e}^{\pm iS^{\frac 32 }\le( \frac 1  3 z^3  +  \le(2  + \frac {\delta_k+\delta_\ell}S \ri)z\ri)} e_{k,\ell} \ot \s_\pm\ ,\ \ \ z \in \gamma_\pm ^{(k,\ell)} 
\ee
is close to the identity jump in any $L^p(\gamma_{\pm}^{(k,\ell)})$, $1\leq p \leq \infty$ with the supremum norm given by 
\be
\| \1  - M^{(k,\ell)}\|_\infty = |c_{k\ell}| {\rm e}^{-\frac 2 3 \le(2S + \delta_k + \delta_\ell\ri)^\frac 32} \to 0\ .
\ee
This shows that the RH problem may be solved by iterations; the first iteration for $\Xi$ yields 
\be
\Xi(\sqrt S z) &\& = \1 - \int_{\gamma_+} \frac {\r(\sqrt S w)\ot \s_+ \d w}{(w-z)} - \int_{\gamma_-} \frac {\tilde\r(\sqrt S w)\ot \s_- \d w}{(w-z)} + \mathcal O\le( \frac { {\rm e}^{-\frac 4 3(2S-2m)^\frac 32}}{1+|z|}\ri) = \\
&\& = \1 - \frac i {\lambda}  [c_{k\ell} \Ai(s_k+s_\ell)] \ot \s_+  +  \frac i {\lambda}  [c_{k\ell} \Ai(s_k+s_\ell)] \ot \s_-  +  \mathcal O\le( \frac {\sqrt {S} {\rm e}^{-\frac 4 3(2S-2m)^\frac 32}}{\sqrt S+|\lambda|}\ri) 
\ee
where the notation $[A_{k\ell}]$ stands for a matrix with entries $A_{k\ell}$.
This yields 
\be
 [\b_1]_{k \ell} =  -i \lim_{\lambda\to\infty}\lambda \Xi_{12}(\lambda) = -c_{k\ell} \Ai (s_k+s_\ell) +  \mathcal O\le(\sqrt{S} {\rm e}^{-\frac 4 3(2S-2m)^\frac 32}\ri) 
\ee

\noindent {\bf Poles.}
It follows from Thm. \ref{Xiexists} that under the stated conditions $\b_1$ exists and finite for all $\vec s=\R^r$ and hence cannot have poles.  

\noindent{\bf Symmetry.} If $C=C^\dagger$ then  (for $\vec s\in \R^r$) $\ov{\r^T(\ov \lambda)}  = -\r(-\lambda)$ (see (\ref{ncHMcL})) and the jump matrices $M(\lambda)$ then satisfy 
\be
M^\dagger (\ov \lambda) = \wh \s_3 M^{-1} (-\lambda) \wh \s_3\ ,\qquad
M(\lambda) = \1_{2r}  + {\rm e}^{\theta(\lambda)} C {\rm e}^{\theta(\lambda)} \ot \le(\s_+\chi_{_{\gamma_+}}(\lambda)+ \s_-\chi_{_{\gamma_-}}(\lambda)\ri)\ .
\ee
The contours of jump satisfy (also) $\ov \gamma_+  = \gamma_-$.  Then $\Xi^{-\dagger}(\ov \lambda):= \ov{\Xi^{-T}(\ov \lambda)} = \wh \s_3 \Xi(\lambda) \wh \s_3$. This implies ($\s_3\s_2 \s_3 = -\s_2$)
\be
\Xi(\lambda) = \1 + \frac {\a_1\ot \s_3 + \b_1 \ot \s_2} \lambda + \dots\  \ \Rightarrow \ \ 
\Xi^{-1}(\lambda) = \1 - \frac {\a_1 \ot \s_3 + \b_1 \ot \s_2}{\lambda}  + \dots\\
\Xi^{-\dagger}(\ov \lambda) = \1 - \frac {\a_1^\dagger   \ot \s_3 + \b_1^\dagger \ot \s_2}{\lambda}  + \dots\ \ \Rightarrow\ \ 
\wh \s_3 \Xi^{-\dagger}(\ov \lambda) \wh \s_3 =  \1 + \frac {-\a_1^\dagger  \ot \s_3 + \b_1^\dagger \ot \s_2}{\lambda}  + \dots
\ee 
which shows immediately $\b_1 = \b_1^\dagger$ (as well as $\a_1 = -\a_1^\dagger$). 

\noindent{\bf Formula for the determinant.}
From Corollary (\ref{corsimple}) we deduce that $\D \ln \det (\Id - \mathcal Ai_{\vec s}^2) = -2i\Tr \a_1$ and together with  eq. (\ref{alphabeta}) we have 
\be
\D^2 \ln \det (\Id - \mathcal Ai_{\vec s}^2) = -4 \Tr (\b_1^2)
\ee
from which the formula follows immediately by integration as in the usual Tracy--Widom distribution.
 \QED

\subsection{Noncommutative PXXXIV}
Similarly we can prove
\begin{shaded} 
\begin{theorem}
\label{thm62}
Let $a_1,a_2$ be the coefficient matrices in the expansion of the  solution $\Gamma$ of Problem \ref{prob1}  as in formula (\ref{314}) 
and define $\Phi(\lambda):= \Gamma(\lambda) {\rm e}^{\theta(\lambda)\ot \s_3}$.
Then 
\be
\pa_{s_j} \Phi &\& = V_j\Phi\ ,\ \ \ \pa_\lambda \Phi = B\Phi\\
V_j&\& := \lambda^2 {\mathrm e}_{j}\ot \s_- +  
 i[a_1,{\mathrm e}_{j}]\ot \1 
    -2  \{b_2 ,{\mathrm e}_{j}\}   \ot \s_- 
- {\mathrm e}_{j}\ot \s_+ \ .
\ee
\be
B(\lambda)&\&  =\!\! \le[
\begin{array}{cc}
0 &\hspace{-10pt} -\frac \lambda 2 \\
\frac {\lambda^3} 2\! +\! \lambda\le(\ss\! -\!\frac i 2 a_1'\ri)  &\hspace{-10pt} 0 
\end{array}
\ri]  \!+\!  \frac 1 \lambda \!\le[
\begin{array}{cc}
 i[a_1,\ss] - \frac i 4 a_1''
  &  - \ss -\frac i 2 a_1'\\[6pt]
\!\!\! 2ia_1\!+ \!2[a_2,\ss]\! +\! 2[\ss,a_1]a_1 \!-\! \frac 1 2(a_1')^2
 & 
\!\! \1\! +\! i[a_1,\ss]\! +\! \frac i 4 a_1''
\!\! \!\end{array}
\ri]
\label{LaxBfinal}
\ee
Denoting by $\D = \sum_{j=1}^r \pa_{s_j}$, so that $\D \ss = \1$, we have 
\be
\D\Phi = V_\D \Phi\ ,\ \ \ V_\D = \lambda^2 \1\ot \s_- -2i \D a_1 \ot \s_-  - \1 \ot \s_+ = \le[
\begin{array}{cc}
0 & -\1\\
\lambda^2 - 2i\D a_1 & 0 
\end{array}
\ri]
\ee
The matrices $a_1,a_2$ satisfy the equations (prime denotes action of $\D$) 
\be
&\& a_1''' =  8i [a_1,\ss]a_1 + 8a_1 + 8i[\ss,a_2] + 6i (a_1')^2 + 4 \{a_1',\ss\} \label{tiso1} \\
&\& a_2' = a_1' a_1\label{tiso2}
\ee
\end{theorem}
\end{shaded}
\br
\label{rem61}
Differentiating (\ref{tiso1}) and using (\ref{tiso2})  one obtains an ODE for the matrix $a_1$ 
\be
a_1^{iv} = 6i\{a''_1, a_1'\} + 8ia_1' [ \ss,a_1] + 8i [a_1, \s a_1'] + 8i \ss [a_1',a_1] + 4 \{\ss , a_1''\}+ 16 a_1'
\ee
\er
\br
\label{rem62}
If $r=1$, then we are in the commutative setting  and $\ss = s$ is just a scalar. Then the term involving $a_2$ in (\ref{tiso1})  drops out and we obtain ($'=\pa_s$) 
\be
\label{P34}
a_1''' = 8a_1 + 6i(a_1')^2 + 8 s\, a_1'
\ee
If we take the equation (\ref{P34}) and we differentiate once we obtain, for $a_1'$, the equation (\ref{P34int}) (up to rescaling). For this reason we will call the system (\ref{tiso1}, \ref{tiso2}) the {\bf noncommutative Painlev\'e\ XXXIV equation}.
\er
\begin{cor}
\label{CorTW1}
Denoting by $F_{1}^{(nc)} (\vec s)$ the  Fredholm determinant of the operator $\Id + {\mathcal Ai}(\bullet;\vec s)$ on $L^2(\R_+, \C^r)$ we have 
\be
\pa_{s_j} F_1^{(nc)}(\vec s) = i (a_1)_{jj}
\ee 
and $a_1$ is a solution of the noncommutative PXXXIV equation (\ref{tiso1}, \ref{tiso2}). In particular 
\be
\D F_1^{(nc)}(\vec s) = i \Tr a_1. 
\ee
\end{cor}
{\bf Proof of Cor. \ref{CorTW1}.} 
The Fredholm determinant equals the determinant of $\Id + \K$ as explained already in the proof of Thm. \ref{Xiexists}. Then the formulas above follow simply from Thm. \ref{4.2} and Corollary \ref{corsimple}.
\QED
Before concluding the paper with the proof of Thm. \ref{thm62} we point out that --in a sense-- all the relevant information is already contained in Thm. \ref{thmncHML} because of the matrix Miura relation 
\be
a_1(\vec s) \mathop{=}^{\hbox{\tiny Prop. \ref{RHPrelation}  }} \a_1(\vec s)-i\b_1(\vec s) \mathop{=}^{\hbox{\tiny eq. (\ref{alphabeta})}} -2i  \int_{S}^\infty\!\!\!\!\! \b_1^2( t+\delta_1, \dots, t+\delta_r)\d t - i \b_1 (\vec s) 
\ee
where $\vec s= (S+\delta_1,\dots S+\delta_r)$ and $\b_1$ is the noncommutative Hastings--McLeod family of solutions (depending on $C$) in Theorem \ref{thmncHML}.
This immediately yields the 
\begin{cor}
\label{CorTW2}
The Fredholm determinant of the matrix Airy convolution kernel $\mathcal Ai_{\vec s}$ satisfies
\be
\det\le(Id + \mathcal Ai_{\vec s}\ri) = \exp\le[  \int_S^\infty  \Tr \le(\beta_1(t+\delta_1,\dots, t+\delta_r) + 2(t-S)\b_1^2(t+\delta_1, \dots, t+\delta_r)  \ri)\d t\ri]
\ee
where $\b_1(\vec s)$ is the Hastings-McLeod family of solutions to noncommutative Painlev\'e\ II as in Thm. \ref{thmncHML}.
\end{cor}

\noindent {\bf Proof of Thm. \ref{thm62}.}
Recall that $\L = \1 \ot L$ and 
\be
L\s_3 L^{-1}= \frac i \lambda \s_+ -i \lambda \s_-\ ,\ \ \ 
L \s_2 L^{-1} = \frac 1  \lambda \s_+ + \lambda \s_-\ ,\ \ \ 
L\s_1 L^{-1}= \s_3 
\ee
\be
\pa_j \Gamma(\lambda) + i \lambda  \Gamma(\lambda) {\mathrm e}_{j} \ot  \s_3 = \le(V^{(2)}_j \lambda^2 +V^{(1)}_j \lambda+ V^{(0)}_j \ri) \Xi(\lambda )\nonumber \\
\L(\lambda) \frac {\pa_j(a_1\ot \s_3 ) }\lambda  +\dots  + i \lambda  \L (\lambda) \le(\1 + \frac {a_1\ot \s_3 + b_1\ot \s_1 }\lambda + \frac {\a_2\ot \1 + b_2\ot \s_1}{\lambda^2}+\dots \ri){\mathrm e}_{j}\ot \s_3 =\nonumber \\=
 \le(V^{(2)}_j \lambda^2 +V_j^{(1)} \lambda + V_j^{(0)} \ri)\L\le(\1 + \frac {a_1\ot \s_3 }\lambda + \frac {a_2\ot \1 + b_2\ot \s_1}{\lambda^2}+\dots \ri)\label{917}
\ee 
\be
-i\pa_j a_1\ot\s_-  + \frac i {\lambda ^2} \pa_j a_1\ot \s_+  + \frac {a_2 \ot \1 + b_2\ot \s_3}{\lambda^2} +\dots  +\\
+ i \lambda   \le(\1 - i a_1\ot \s_- + \frac i{\lambda^2} a_1\ot \s_+ +  \frac {a_2\ot \1 + b_2\ot \s_3}{\lambda^2}+\dots \ri){\mathrm e}_{j}\ot \le(\frac i \lambda \s_+ - i \lambda\s_-\ri) =
\nonumber \\
 \le(V^{(2)}_j \lambda^2 +V_j^{(1)} \lambda + V_j^{(0)} \ri)
  \le(\1 - i a_1\ot \s_- + \frac i{\lambda^2} a_1\ot \s_+ +  \frac {a_2\ot \1 + b_2\ot \s_3}{\lambda^2}+\dots \ri)
\ee 
\be
= \lambda^2(\1 - ia_1\ot \s_-)  {\mathrm e}_{j}\ot \s_-  + \le(ia_1\ot \s_+ + a_2\ot \1 + b_2 \ot \s_3 \ri) {\mathrm e}_{j}\ot \s_- 
-(\1-ia_1\ot \s_-) {\mathrm e}_{j}\ot \s_+
\nonumber =\\
=\lambda^2 V^{(2)}_j (\1 - ia_1\ot \s_-) + V_j^{(2)} \le( i a_1\ot \s_+ + a_2 \ot \1 + b_2\ot \s_3 \ri) + V_j^{(0)} (\1 - ia_1\ot \s_-)
\ee

We thus have 
\be
V^{(2)}_j (\1 - ia_1\ot \s_-)  = (\1 - ia_1\ot \s_-)  {\mathrm e}_{j}\ot \s_- \ \ \Rightarrow \ \ V_j^{(2)} = {\mathrm e}_{j}\ot \s_-
\ee
\be
 V_j^{(0)} &\&= 
 i[a_1,{\mathrm e}_{j}]\ot \1 
  +\big( [a_2,{\mathrm e}_{j}]      -  \{b_2 ,{\mathrm e}_{j}\} -  [ a_1, {\mathrm e}_{j}]a_1 - i\pa_j a_1 \big)    \ot \s_- 
- {\mathrm e}_{j}\ot \s_+  
\ee
Looking at the $\lambda^{-1}$ coefficient one finds the following identity 
\be
-i \pa_j a_1 = [{\mathrm e}_{j},a_2] + [a_1,{\mathrm e}_{j}]a_1 -\{{\mathrm e}_{j}, b_2\} \label{a1j}
\ee
which allows us to rewrite 
\be
 V_j^{(0)} &\&= 
 i[a_1,{\mathrm e}_{j}]\ot \1 
    -2  \{b_2 ,{\mathrm e}_{j}\}   \ot \s_- 
- {\mathrm e}_{j}\ot \s_+  
\ee
Summing up (\ref{a1j}) for $j=1,\dots r$ we also have 
\be
 \D a_1= -2i b_2 \ .
\label{653}
\ee
We will need also more information from (\ref{917}) by looking at the coefficients of the negative powers of $\lambda$ in particular we will need this for $V_\D:= \sum_{j=1}^{r}$.
 To do so it is convenient to multiply (\ref{917}) on the left by $\L^{-1}$. 
Below we list the results of lengthy but completely straightforward inspections. We list the entry of the coefficient of $\lambda^{j}$ in the form $[\lambda^{j}]_{k,\ell}$. 
\be
\begin{array}{c|c}
[\lambda^{-1}]_{1,2} & 
b_2 = \frac i 2 a_1'\\
\hline\\
 \ds [\lambda^{-2}]_{1,2}  & 
\ds b_3 = -\frac 1 2 a_1' a_1 - \frac i 4 a_1''\\
\hline\\
\ds  [\lambda^{-2}]_{1,1}  & 
\ds a_2' =  \frac 1 2 a_1' a_1 - i b_2a_1 = a_1' a_1  \\
\hline\\
\ds [\lambda^{-3}]_{1,2}  & b_4 =
\frac i 4 a_1' b_2 + \frac 1 2 b_2^2 + \frac i 4 a_1' a_2 + \frac 1 2 b_3' + \frac 1 2 b_2 a_2 = 
 -\frac 12 (a_1')^2 + \frac i 2 a_1' a_2 - \frac 14 a_1'' a_1 - \frac i 8 a_1''' 
\end{array}
\label{656}
\ee

A similar and completely straightforward computation (involving longer algebra) yields 
\be
B(\lambda)&\&  = \le[
\begin{array}{cc}
0 & -\frac \lambda 2 \\
\frac {\lambda^3} 2 + \lambda\le(\ss -b_2\ri)  & 0 
\end{array}
\ri] +\\
&\& +  \frac 1 \lambda \le[
\begin{array}{cc}
b_3 - i b_2 a_1+ i[a_1,\ss] &  - \ss - b_2\\
ia_1 - b_4 -\{\ss,b_2\}  - ib_3 a_1 +[a_2,\ss] + [\ss,a_1]a_1 + b_2(b_2 - a_1^2 +  a_2) & 
i[a_1,\ss] + ib_2 a_1 - b_3 + \1 
\end{array}
\ri]
\ee
The expression can be simplified using (\ref{656}) to give
\be
B(\lambda)&\&  = \le[
\begin{array}{cc}
0 & -\frac \lambda 2 \\
\frac {\lambda^3} 2 + \lambda\le(\ss -\frac i 2 a_1'\ri)  & 0 
\end{array}
\ri] +\\
&\& +  \frac 1 \lambda \le[
\begin{array}{cc}
 i[a_1,\ss] - \frac i 4 a_1''
  &  - \ss -\frac i 2 a_1'\\
ia_1 +[\ss,a_2]  - \frac i 2 \{\ss, a_1'\}  + [x,a_1]a_1+ \frac 1 4 (a_1')^2 + \frac i 8 a_1'''
 & 
\1 + i[a_1,\ss] + \frac i 4 a_1''
\end{array}
\ri]
\label{LaxB}
\ee

One then has to write the zero-curvature equations 
\be
\pa_\lambda V_\D - \D B + [V_\D , B]=0\label{zerobv}
\ee

which yield
\be
&\& a_1''' =  8i [a_1,\ss]a_1 + 8a_1 + 8i[\ss,a_2] + 6i (a_1')^2 + 4 \{a_1',\ss\} \label{iso1} \\
&\& a_2' = a_1' a_1\label{iso2}
\ee
where the first equation comes from the entry $(1,1)$ of the coefficient in $\lambda^{-1}$ of (\ref{zerobv}), while the  second equation comes from (\ref{656}); all other entries of (\ref{zerobv}) are then automatically zero.
One can use (\ref{iso1}) to simplify further the expression for $B$ as given in the statement of the theorem.
\QED
\def\cprime{$'$}

\end{document}